\documentclass[conference]{IEEEtran}
\usepackage{etoolbox}
\makeatletter
\patchcmd{\@makecaption}
  {\\}
  {.\ }
  {}
  {}
\makeatother
\usepackage{booktabs}
\usepackage{tikz}
\usepackage{url}
\usepackage{pdfpages}
\usepackage{afterpage}
\usepackage{color, colortbl}
\usepackage{graphicx}
\usepackage{amsmath}
\usepackage{subfigure}
\usepackage{footmisc}

\usepackage{algorithm}
\usepackage[noend]{algpseudocode}
\usepackage{varwidth}
\usepackage[flushleft]{threeparttable}
\usepackage{hyperref}

\definecolor{Gray}{gray}{0.9}
\definecolor{LightCyan}{rgb}{0.88,1,1}
\pdfoutput=1
\graphicspath {{figures/}}

\begin{document}

\title{Supervised Learning Based Algorithm Selection for Deep Neural Networks}

\author{\IEEEauthorblockN{Shaohuai Shi, Pengfei Xu, Xiaowen Chu}
\IEEEauthorblockA{Department of Computer Science, Hong Kong Baptist University
\\\{csshshi, pengfeixu, chxw\}@comp.hkbu.edu.hk}
}

\maketitle

\begin{abstract}
Many recent deep learning platforms rely on third-party libraries (such as cuBLAS) to utilize the computing power of modern hardware accelerators (such as GPUs). However, we observe that they may achieve suboptimal performance because the library functions are not used appropriately. In this paper, we target at optimizing the operations of multiplying a matrix with the transpose of another matrix (referred to as NT operation hereafter), which contribute about half of the training time of fully connected deep neural networks. Rather than directly calling the library function, we propose a supervised learning based algorithm selection approach named MTNN, which uses a gradient boosted decision tree to select one from two alternative NT implementations intelligently: (1) calling the cuBLAS library function; (2) calling our proposed algorithm TNN that uses an efficient out-of-place matrix transpose. We evaluate the performance of MTNN on two modern GPUs: NVIDIA GTX 1080 and NVIDIA Titan X Pascal. MTNN can achieve 96\% of prediction accuracy with very low computational overhead, which results in an average of 54\% performance improvement on a range of NT operations. To further evaluate the impact of MTNN on the training process of deep neural networks, we have integrated MTNN into a popular deep learning platform Caffe. Our experimental results show that the revised Caffe can outperform the original one by an average of 28\%. Both MTNN and the revised Caffe are open-source.
\end{abstract}

\begin{IEEEkeywords}
Linear Algebra; Matrix Multiplication; Transpose; GPU; Deep Neural Networks
\end{IEEEkeywords}

\IEEEpeerreviewmaketitle

\section{Introduction} \label{introduction}
Deep neural networks have recently achieved great success in computer vision, speech recognition, and natural language processing \cite{lecun2015lenet}\cite{krizhevsky2012imagenet}. The forwarding and backwarding phases in the backpropagation based training process of a deep neural network requires two different forms of matrix multiplication (i.e., Equation \ref{equation:1} and Equation \ref{equation:2}), which dominate the training time. The regular form of matrix multiplication for two row-major matrices \textit{A} and \textit{B} can be represented as follows:
\begin{equation}
\label{equation:1}
{\textit{C}=\textit{A}\times \textit{B}}
\end{equation}
where $\textit{A}\in R^{m\times k}$, $\textit{B}\in R^{k\times n}$ and $\textit{C}\in R^{m\times n}$. In this paper we call Equation \ref{equation:1} NN operation (N means no transpose). There is another form of matrix multiplication: \textit{A} multiplied with the transpose of \textit{B}, i.e.,
\begin{equation}
\label{equation:2}
{\textit{C}=\textit{A}\times \textit{B}^T}
\end{equation}
where $\textit{B}^T$ is the transpose of \textit{B}, $B^T_{ji} = B_{ij}$ and $\textit{B}\in R^{n\times k}$. In this paper we call Equation \ref{equation:2} NT operation (T means transpose).

The time complexity of schoolbook matrix multiplication is $O(m\times k\times n)$, which makes it very time-consuming for large matrices. Nowadays, there exist many optimized software libraries for matrix operations, including ATLAS, LAPACK, OpenBLAS, GotoBLAS, Intel MKL, Eigen, cuBLAS, etc. As GPUs have become mainstream hardware accelerators, the cuBLAS library from NVIDIA becomes a major linear algebra library for state-of-the-art deep learning software tools \cite{shi2016benchmarking}. For example, the SGEMM function in cuBLAS library running on an NVIDIA K40M card can achieve about 3000 GFLOPS when performing single-precision floating-point matrix multiplication, which is up to 17x faster than the MKL library on Intel CPU IvyBridge E5-2697v2 @ 2.70GHz \cite{cublasperformance}.

Some recent work has been proposed to understand and improve the performance of NN operations on GPUs \cite{lai2013performance}. Considering the complexity of GPU architectures, it is very challenging to design a single algorithm or a single set of kernel configuration that is optimal for all cases; hence autotuning method has become an attractive approach to choosing the best algorithms or kernel configurations for GPUs \cite{kurzak2012autotuning}\cite{abdelfattah2016performance}. However, the NT operations have not received much attention from the research community. Our previous work shows that many state-of-the-art deep learning software tools overlook the importance of NT operations and only achieve suboptimal performance for some deep neural networks \cite{shi2016benchmarking}. In this paper, we first show that the performance of NT operations by cuBLAS is often much lower than that of NN operation on recent GPUs. We then propose a simple method called TNN which implements the NT operation by carrying out efficient out-of-place matrix transpose first and then performing an NN operation. In general, TNN outperforms cuBLAS for large matrices, but it is not as efficient as cuBLAS for small matrices. In order to achieve the best average performance, we design an algorithm selection method named MTNN, which can intelligently select the appropriate algorithm to carry out the NT operations based on some GPU architecture information and matrix sizes. Notice that the idea of algorithm selection dates back to 1976 \cite{rice1976algorithm} and becomes very successful in recent years to choose optimal implementation from a set of algorithms \cite{spillinger2015matrix}\cite{benatia2016sparse}\cite{sedaghati2015automatic}. In order to verity the effectiveness of MTNN, we integrate it into a popular real world deep learning platform Caffe \cite{jia2014caffe} which relies on cuBLAS to accelerate its NN and NT operations on GPUs. We evaluate the performance of MTNN and the revised Caffe on two modern GPUs: NVIDIA GeForce GTX1080 and Titan X Pascal. The experimental results show that (1) our MTNN solution achieves up to  54.03\% improvement on average over the NT operation of cuBLAS; and (2) the revised Caffe\footnote{Our source codes can be found here: \url{https://github.com/hclhkbu/caffe-optimized}} achieves 28\% speedup over the original Caffe on the tested GPUs.

The rest of the paper is organized as follows. We present the motivation of this work in Section \ref{motivation}, and then introduce the related work in Section \ref{relatedwork}. The TNN method is described in Section \ref{naivemethod}, followed by our MTNN framework in Section \ref{modelbasedmethod}. Experimental results are presented in Section \ref{results}. We conclude the paper and discuss our future work in Section \ref{conclusion}.

\section{Motivation} \label{motivation}
On deep neural networks, especially the fully connected networks \cite{hecht1989theory}, matrix-matrix multiplication (i.e., NN operations) and matrix-matrix-transpose multiplication (i.e., NT operations) are the two major computational tasks for the training process. Both types of matrix multiplication are commonly implemented by the SGEMM routine of BLAS library in practice. The standard SGEMM has the following form:
\begin{center}
$C=\alpha \cdot op(A) \times op(B)+\beta \cdot C$
\end{center}
where $op$ represents whether the matrix is transposed or not, and $\alpha$ and $\beta$ are scalars. To simplify the calculation, we ignore the second term and set $\alpha$ to 1. In cuBLAS, the SGEMM API is ``cublasSgemm'', in which the second and the third parameters are the values of $op$ for \textit{A} and \textit{B} respectively. The value of $op$ can be ``CUBLAS\_OP\_T'' (transpose) or ``CUBLAS\_OP\_N'' (no transpose). To understand the performance difference between NN and NT operations in cuBLAS, we conduct experiments to evaluate the running time performance of SGEMM for NN and NT operations with different sizes of input matrices. Table \ref{table:hardwaresetup} shows the details of our two tested platforms.

\begin{table}[!ht]
\centering
\caption{The experimental GPU hardware with CUDA-8.0}
\label{table:hardwaresetup}
\begin{tabular}{|l|c|c|c|c|}
\hline
GPU Model  &  Cores  &   Memory  & OS & Core frequency\\\cline{1-5}
\hline
\hline
GTX1080 & 2560 & 8 GB & Ubuntu 14.04 & 1607 MHz \\\cline{1-5}
Titan X & 3584 & 10 GB & Ubuntu 14.04 & 1417 MHz \\\cline{1-5}
\end{tabular}
\end{table}

\begin{figure}[!htbp]
  \centering
  \subfigure[GTX1080]
  {
    \includegraphics[width=0.45\linewidth]{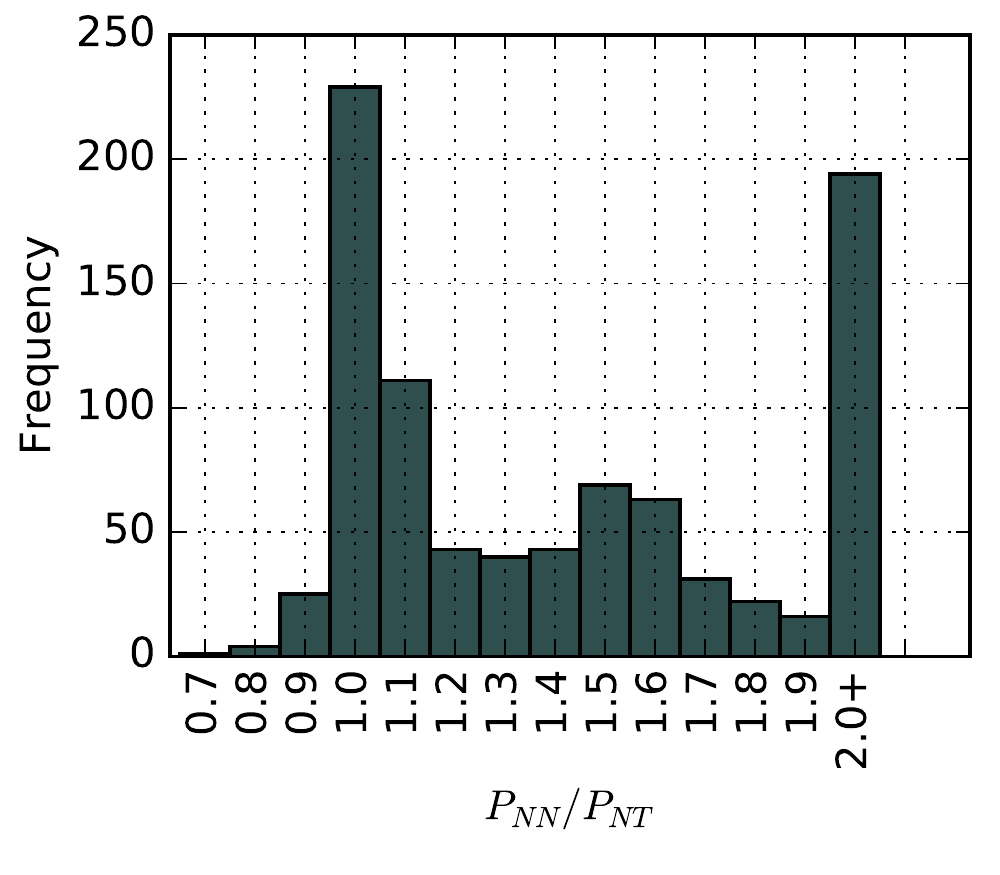}
    \label{fig:rawntvsnn1080}
  }
  \subfigure[TitanX]
  {
    \includegraphics[width=0.45\linewidth]{rawntvsnn1080_frequency.pdf}
    \label{fig:rawntvsnntitanx}
  }
\caption{The frequency of performance ratio of $P_{NN}$ over $P_{NT}$ among 1000 tested cases on each GPU. The last value (i.e., 2.0+) of x-axis means: $P_{NN}/P_{NT} \geq 2.0 $.}
\label{fig:rawntvsnn}
\end{figure}

We use $P_{algorithm}$ to denote the performance of a specific $algorithm$ with the unit of GFLOPS. To illustrate the difference between $P_{NN}$ and $P_{NT}$, we run experiments for 1000 cases with different matrix sizes and show the distribution of resulted $P_{NN}/P_{NT}$ in Fig. \ref{fig:rawntvsnn}. It is noted that, in most cases, the performance of NN ($P_{NN}$) is much better than that of NT ($P_{NT}$) because there is no overhead of matrix transpose. The percentages of the number of cases that $P_{NN}$ is higher than $P_{NT}$ are 71\% and 62\% on GTX1080 and Titan X respectively. More surprisingly, there are around 20\% of cases with $P_{NN}/P_{NT} \geq 2.0$ on both GPUs. The low performance of NT of cuBLAS may be caused by the inefficient memory access to the elements of \textit{B}. Another possible reason is that cuBLAS uses the slow in-place matrix transpose algorithm to reduce the memory footprint \cite{gomez2016place}. Observing this low efficiency issue, we are motivated to propose a method (TNN) for NT operations which finds the transpose of \textit{B} first and then calls NN function of cuBLAS to finish the calculation of $A\times B^T$ on GPUs. The performance of TNN is better than cuBLAS in most cases, but still there exist cases that cuBLAS outperforms our TNN. To this end, we further design an algorithm selection approach to select an appropriate algorithm from the set \{TNN, NT\} based on a supervised learning algorithm. Notice that TNN requires that the GPU memory is large enough to store the additional $B^T$. If that is not the case, our framework will simply choose the original NT operations.

\section{Related Work} \label{relatedwork}
SGEMM algorithm in cuBLAS has been intensively optimized on GPUs by kernel optimizations \cite{lai2013performance}\cite{volkov2008benchmarking}\cite{nath2010improved}\cite{tan2011fast} and auto-tuning algorithms \cite{li2009note}\cite{kurzak2012autotuning}\cite{abdelfattah2016performance}. The information of different levels of GPU memory access latency \cite{volkov2008benchmarking} and instruction computation \cite{lai2013performance} are extracted to help increase the parallelism of GPU kernels, which can achieve excellent performance that is close to the theoretical hardware capacity based on the block-based matrix-matrix multiplication algorithm. Targeting at Fermi GPU of DGEMM (GEMM in double precision), R. Nath et al. \cite{nath2010improved} propose a double blocking algorithm to reduce the impact of latency in accessing registers and the shared memory, which can achieve up to 58\% of the peak performance. Even though there is a well-designed kernel on GPU, the discrepancy among distinct GPUs could require different configurations to obtain best performance. Instead of conducting detailed kernel analysis, auto-tuning methods have been investigated to select the optimal configuration to achieve better performance of the kernel \cite{li2009note}\cite{kurzak2012autotuning}\cite{abdelfattah2016performance}.

However, little work has been done to evaluate the performance of the NT operations. Since $B_{ji}^T=B_{ij}$, we can perform NT by changing the access of a row to the corresponding column of matrix \textit{B} with SGEMM routine. However, it might cause extra latencies due to uncoalesced global memory access and conflicted shared memory access when fetching the column elements of matrix \textit{B}. The kernel optimization of NT is challenging because its performance depends not only on the GPU architecture, but also on the input matrix size. Therefore, instead of optimizing the kernel algorithm, we first propose a simple approach called TNN as an alternative to SGEMM. We notice that TNN can significantly outperform SGEMM in many cases, but sometimes its performance could be worse than SGEMM. To this end, we formulate an algorithm selection problem in order to select the appropriate algorithm for each NT operation. 

Machine learning approaches become useful in choosing more efficient algorithms with high accuracy \cite{spillinger2015matrix}\cite{sedaghati2015automatic}\cite{benatia2016sparse}. Spillinger et al. \cite{spillinger2015matrix} exploit SVM model \cite{cortes1995support} to predict the better implementation of matrix multiplication algorithm at runtime among two implementations of MKL and CARMA on three different CPU platforms, which achieves about 26\% performance improvement on average. Beside the SVM models which have been applied to solve algorithm selection problems \cite{spillinger2015matrix}\cite{sedaghati2015automatic}, the decision tree classifier is also used to solve the automatic selection of sparse matrix representation on GPUs and it obtains no more than 1.05x average slowdown compared to the existing ideal approach \cite{sedaghati2015automatic}. In this paper, we make use of machine learning techniques to choose the more efficient algorithm between our proposed TNN and the original cuBLAS implementation to improve the average performance in calculating $C=A\times B^T$.

\section{TNN: transpose before multiply} \label{naivemethod}
As we already show in Fig. \ref{fig:rawntvsnn}, directly calculating $C=A\times B^T$ is usually inefficient. We propose a simple TNN method which replaces the one-step NT operation by two-step operation, i.e., transpose \textit{B} first and then make use of NN. The overall performance can be improved if $T_{TNN} = T_{transpose} + T_{NN} < T_{NT}$, where $T_{algorithm}$ is the computation time of $algorithm$. Note that $T_{transpose}$ includes the time of GPU memory allocation and release.

Matrix transpose is a memory bound operation \cite{ruetsch2009optimizing}. There are two very different ways to perform matrix transpose: in-place and out-of-place. The in-place matrix transpose algorithm does not require extra memory space. However, the in-place matrix transposition can be factored as a product of disjoint circles \cite{hungerford2012abstract}, and the number of circles could be much lower in rectangular matrices and their length is not uniform, which results in the difficulty in parallelization \cite{gomez2016place}. The state-of-the-art implementation of in-place matrix transposition achieves only 51.56 GB/s and 22.74 GB/s on GTX 980 (with a peak memory bandwidth of 224 GB/s) and Telsa K20 (with a peak memory bandwidth of 208 GB/s) respectively with single precision \cite{gomez2016place}.

\begin{figure*}[!ht]
  \centering
  \includegraphics[width=\linewidth]{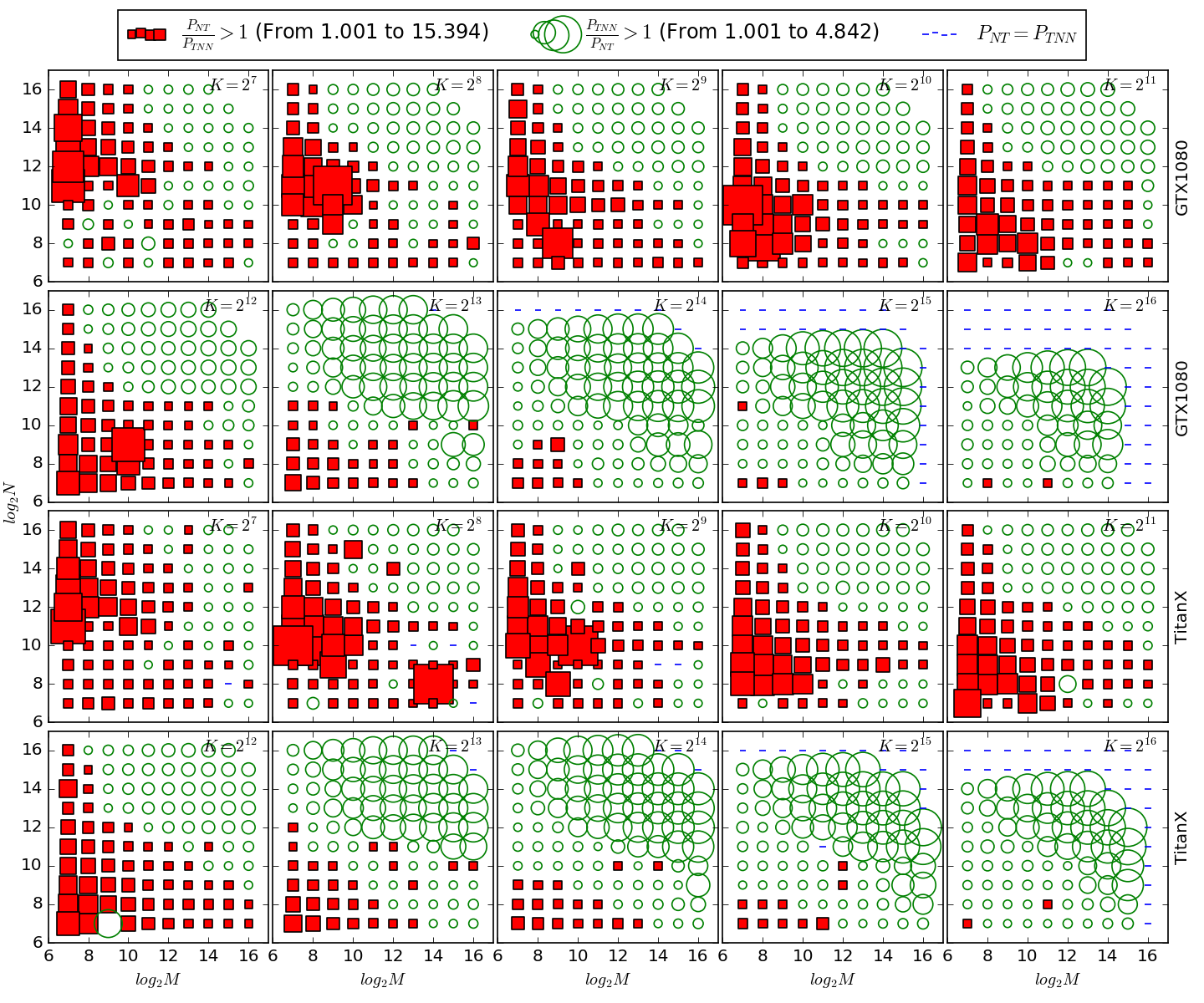}\\
\caption{The performance comparison between NT by cuBLAS and TNN in calculating $C=A\times B^T$. The red rectangle symbol in the legend indicates that the performance of NT is better than TNN; the green circle symbol indicates that the performance of NT is worse than TNN; and the blue dash symbol indicates that the performances of NT and TNN are equal. The size of rectangle and circle symbols reflect the value of $P_{NT}/P_{TNN}$ and $P_{TNN}/P_{NT}$ respectively: a larger symbol size indicates a higher ratio value. The top two rows are the results for GTX1080, and the bottom two rows are the results for Titan X.}
\label{fig:ntvstnn}
\end{figure*}

\begin{figure}[!htbp]
  \centering
  \subfigure[GTX1080]
  {
    \includegraphics[width=0.45\linewidth]{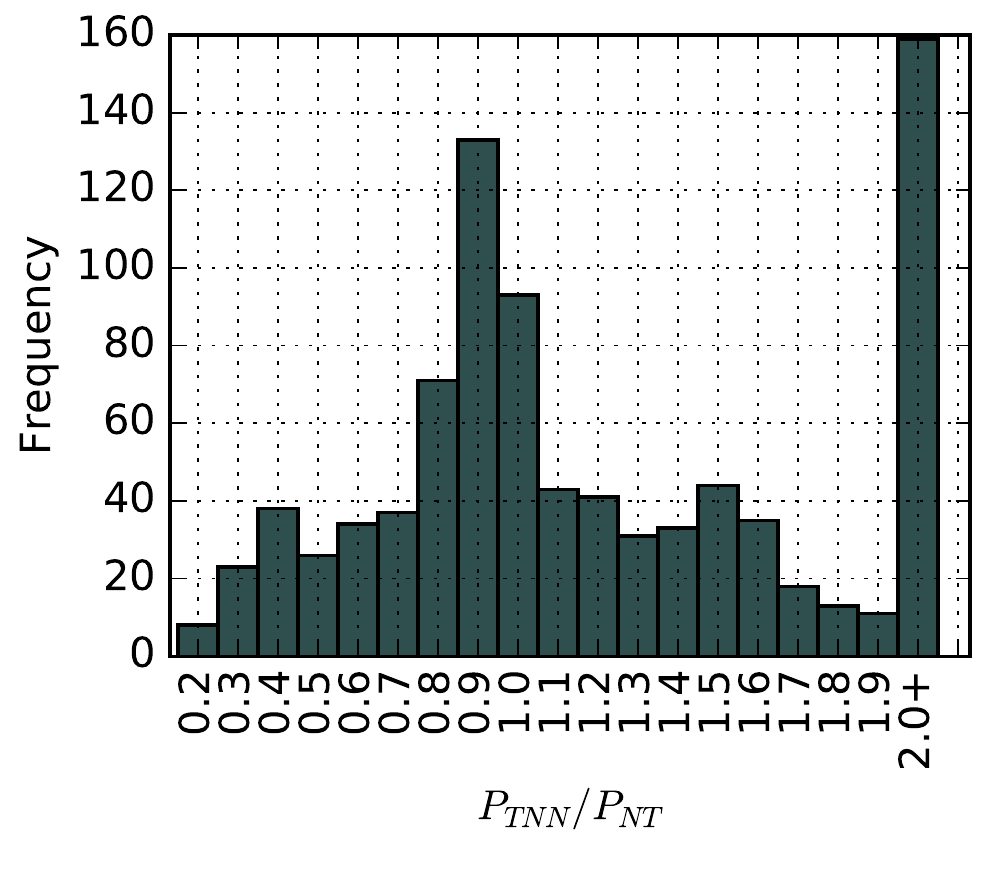}
  }
  \subfigure[TitanX]
  {
    \includegraphics[width=0.45\linewidth]{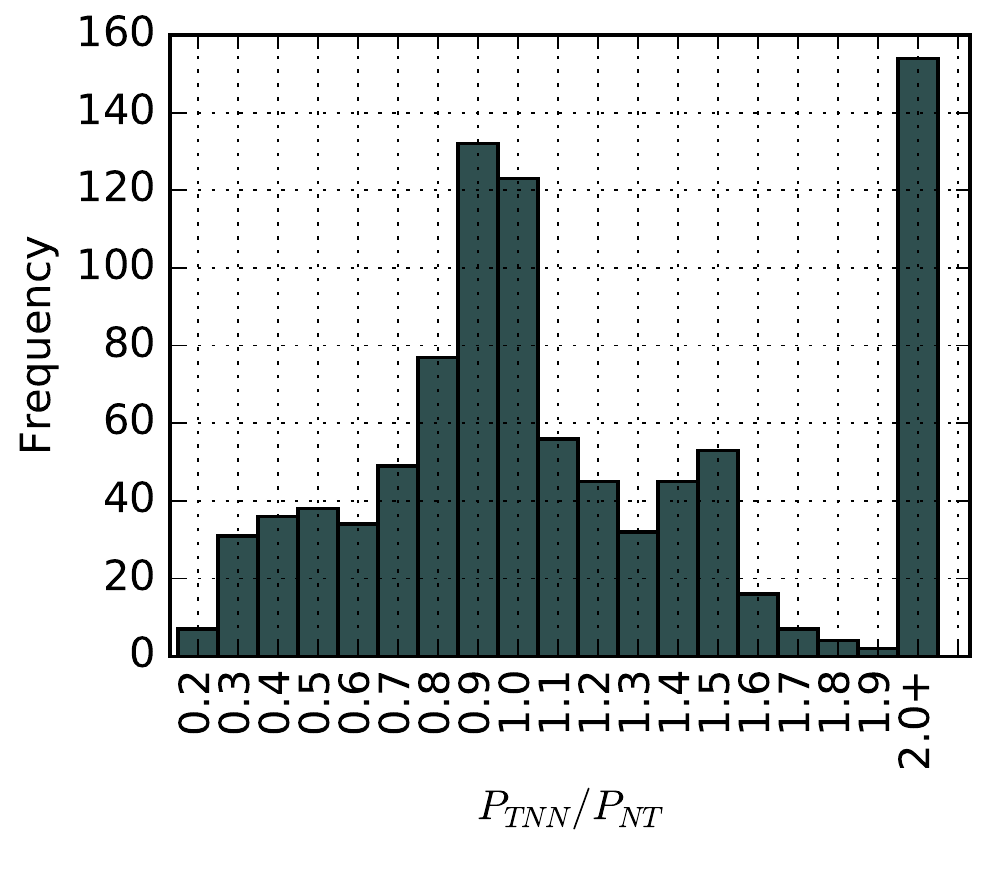}
  }
\caption{The frequency of performance ratio of $P_{TNN}$ over $P_{NT}$ among 1000 tested cases on each GPU. The last value (e.g, 2.0+) of x-axis means: $P_{TNN}/P_{NT}$ is greater than or equal to that value (e.g., $P_{TNN}/P_{NT} \geq 2.0$).}
\label{fig:ntvstnnfreq}
\end{figure}

\begin{algorithm}
\caption{TNN} \label{algo:naive}
\begin{algorithmic}[1]
\Procedure{TNN}{A, B, C, m, n, k}
\footnotesize{
\State // Allocate GPU memory for transpose of B
\State BT = cudaMemAlloc(n*k*sizeof(float))
\State // Tranpose B on GPU and store to BT
\State transposeOnGPU(B, n, k, BT)
\State // Call gemm of cuBLAS with NN parameters
\State \begin{varwidth}[t]{\linewidth}
cublasSgemm(\par
\hskip\algorithmicindent cublasHandler, \par
\hskip\algorithmicindent CUBLAS\_OP\_N, \par
\hskip\algorithmicindent CUBLAS\_OP\_N, \par
\hskip\algorithmicindent A, lda, \par
\hskip\algorithmicindent BT, ldb, \par
\hskip\algorithmicindent C, ldc, ...)
\end{varwidth}
\State // Free GPU memory of BT
\State cudaFree(BT)
\EndProcedure
}
\end{algorithmic}
\end{algorithm}
On the contrary, the out-of-place matrix transposition can exploit the GPU shared memory to achieve an efficient utilization of GPU memory bandwidth. In \cite{ruetsch2009optimizing}, the optimized transpose kernel achieves up to 80\% of peak bandwidth on tested GPUs, which is much higher compared to the in-place algorithm. Therefore, when the rest GPU memory is available to store $\textit{B}^T$ to perform the out-of-place matrix transpose, we can choose the out-of-place transpose routine to implement our TNN algorithm. The pseudo-code of TNN is shown in Algorithm \ref{algo:naive}. Since TNN requires the additional transpose operation on GPU, the time used by transpose operation ($T_{transpose}(n, k)$) should not be larger than difference between NT ($T_{NT}(m, n, k)$) and NN ($T_{NN}(m, n, k)$). In other words, to guarantee $T_{TNN}(m, n, k)$ is smaller than $T_{NT}(m, n, k)$), we have:
\begin{equation}
\label{equation:4}
T_{transpose}(n, k) < T_{NT}(m, n, k) - T_{NN}(m, n, k)
\end{equation}
However, the performance of transpose operation is highly affected by the hardware platform and the size of the matrix. It is difficult to guarantee Equation \ref{equation:4} in practice because there do exist cases that the difference between $T_{NT}(m, n, k)$) and $T_{NN}(m, n, k)$ is small or even $T_{NT} < T_{NN}$, like the cases of $P_{NN}/P_{NT}=1.1$. We show the experimental results of NT and TNN in Fig. \ref{fig:ntvstnn} and Fig. \ref{fig:ntvstnnfreq}. $M$ is the height of matrix \textit{A}, $N$ is the height of matrix \textit{B}, and $K$ is the width of \textit{A} and \textit{B}. In Fig. \ref{fig:ntvstnn}, both x-axis and y-axis are using $log_2$ scale, i.e., the value of $M$ and $N$ are varied from $2^7$ to $2^{16}$. The value of $K$ is also chosen from $2^7$ to $2^{16}$, which forms a total of 1000 cases. To show the detailed visual results, we display all values of $K$ in Fig. \ref{fig:ntvstnn} with various values of $M$ and $N$. In this figure, the red rectangle indicates that the performance of NT is better than TNN; the green circle symbol indicates that the performance of NT is worse than TNN; and the blue dash symbol indicates that the performances of NT and TNN are equal. The size of the symbols is determined by the value $P_{NT}/P_{TNN}$ or $P_{TNN}/P_{NT}$: a larger symbol size indicates a higher value of the ratio.

From Fig. \ref{fig:ntvstnn}, it is noticed that there are some cases that NT outperforms the TNN method, especially when the value of $K$ is small (e.g., there are up to half of the cases that NT is better than TNN when $K$ is 128 on both GPUs). Among all the tested cases, the maximum speedup of TNN over NT is 4.7x, whilst the maximum speedup of NT over TNN is 15.39x. From Fig. \ref{fig:ntvstnnfreq}, it is easy to see that there is a great portion of cases (about 41.5\% on GTX1080 and 43\% on TitanX) that are located in the left side of $P_{TNN}/P_{NT}=1.0$.

Therefore, to perform faster calculations of $C=A \times B^T$, we should choose the NT algorithm and the TNN algorithm appropriately.

\section{MTNN: a Supervised Learning Based Algorithm Selection Method} \label{modelbasedmethod}
In this section, we first formulate the algorithm selection problem as a classification problem for two given input sizes of matrices and a specific GPU platform. Let the class: $-1$ denote $P_{TNN} > P_{NT}$ and the class: $1$ denote $P_{TNN} \leq P_{NT}$. Given a GPU platform: $G$, the size of matrix \textit{A} ($m \times k$) and the size of matrix \textit{B} ($n \times k)$, there exists a function:
\begin{equation}
\label{equation:mapping}
f: (G, m, n, k) \mapsto \{-1, 1\}
\end{equation}
We need to learn a function $\hat{f}$ such that:
\begin{equation}
\label{equation:minimum}
\hat{f} = argmin\sum_{\substack{(G,m,n,k)\in \Omega}}||\hat{f}(G,m,n,k)-f(G,m,n,k)||
\end{equation}
The learning of function $\hat{f}$ can be regarded as a binary classification problem. There are 4 main steps of our supervised-learning based method MTNN. First, we need to construct the training and testing data sets with proper preprocessing of data by benchmarking the performance of NT and TNN. Second, we learn a decision model (i.e., $\hat{f})$ from training samples with supervised machine learning algorithms. Third, we evaluate the learned model on the testing data set. Lastly, we apply the trained model to predict the better implementation (i.e., NT or TNN) in calculating $C=A \times B^T$.

\subsection{Data Collection} \label{datacollection}
According to the results in Fig. \ref{fig:rawntvsnn}, we choose a range of matrices with sizes in $S=\{2^i|i=7, 8, ..., 16\}$. In other words, for all $m$, $n$ and $k$ ($m \in S, n \in S, k \in S$), which has 1000 combinations, we measure the performances of NT and TNN in calculating $C=A \times B^T$. Let $P_{NT}(m, n, k)$ and $P_{TNN}(m, n, k)$ denote the performance of NT and TNN respectively with two matrices \textit{A} and \textit{B}, where $\textit{A}\in R^{m\times k}$ and $\textit{B}\in R^{n\times k}$. The difference value between $P_{NT}(m,n,k)$ and $P_{TNN}(m,n,k)$ is denoted by $D(m,n,k)$. If $D(m,n,k) \geq 0$, then $label=1$, otherwise $label=-1$. Each record is with the following format:

{\centering
($m$, $n$, $k$), $label$\\
}

For each type of GPU, 1000 cases are tested; but some samples that cannot be fitted into memory are not included into the evaluation. So the number of valid samples on each GPU is less than 1000, and the sample distribution is shown in Table \ref{table:sampledist}.
\begin{table}[!ht]
\centering
\caption{Sample distribution on tested GPUs}
\label{table:sampledist}
\begin{tabular}{|l|c|c|c|c|}
\hline
GPU Model &  GTX1080 & TitanX \\\cline{1-3}
\hline
\hline
\# of $-1$ & 649 & 535 \\\cline{1-3}
\# of $1$ & 242 & 406 \\\cline{1-3}
\# of Samples & 891 & 941 \\\cline{1-3}
\textbf{Total} & \multicolumn{2}{c|}{1832}\\\cline{1-3}
\end{tabular}
\end{table}
Besides the variety of input size of matrices, the GPU platform can also be different. Thus, we need to extract the features to represent different GPUs. The details of tested GPUs are shown in Table \ref{table:charofgpus}, which are used as input features of the GPU platform.

\begin{table}[!ht]
\centering
\caption{Characteristics of tested GPUs}
\label{table:charofgpus}
\begin{tabular}{|l|c|c|c|c|}
\hline
GPU & GTX1080 & TitanX\\\cline{1-3}
\hline
\hline
\textbf{Compute Capability} & 6.1 & 6.1\\\cline{1-3}
\textbf{Global Mem (GB)} & 8 & 10\\\cline{1-3}
\textbf{\# of SMs} & 20 & 28\\\cline{1-3}
\textbf{Core Clock (MHz)} & 1607  & 1417 \\\cline{1-3}
\textbf{Mem Clock (MHz)} & 5005 & 5005\\\cline{1-3}
\textbf{Mem Bus Width} & 256  & 384 \\\cline{1-3}
\textbf{L2 Cache (KB)} & 2048 &3072\\\cline{1-3}
\end{tabular}
\end{table}

Combined with different values of the characteristics of GPU in Table \ref{table:charofgpus}, the input sample \textbf{x} is formed as an 8-dimension (5 dimensions from GPU specification and 3 dimensions from matrices size). The first 5 dimensions are global memory ($gm$), the number of SMs ($sm$), core clock ($cc$), memory bus width ($mbw$) and the size of L2 cache ($l2c$). Note that the feature generation is an $O(1)$ computation, which is crucial to reduce the overhead of using the predictor in runtime. The final format of input sample \textbf{x} is as follows:
\begin{center}
($gm,sm,cc,mbw,l2c,m,n,k$), $label$
\end{center}
We do not need to normalize the input feature by using decision tree learning algorithms. By contrast, each dimension of the input feature should be normalized to the range of (0, 1) when training SVMs.
\subsection{Model Training}
Given the training set: $\textbf{S} = \{\textbf{x} | \textbf{x} = (G, m, n, k)\}$, where $G$ is the feature combination in Table \ref{table:charofgpus},  we want to learn function: $\hat{f}$, where
\[
\hat{f}(\textbf{x}) =
	\begin{cases}
	-1, & P_{NT}(\textbf{x}) < P_{TNN}(\textbf{x})\\
	+1, & P_{NT}(\textbf{x}) \geq P_{TNN}(\textbf{x})
	\end{cases}
\]
If $\hat{f}(\textbf{x})=-1$, then we choose TNN, otherwise we choose NT.

\textbf{Learning Algorithm.} SVM \cite{suykens1999least} is a power tool learning algorithm in solving classification problems. And it has been successfully applied to solve algorithm selection problems related to matrix-matrix multiplication \cite{spillinger2015matrix}\cite{benatia2016sparse}. Another powerful learning algorithm: decision tree (DT) is also prosperously used in solving the problem of automatic best algorithm selection \cite{sedaghati2015automatic}, and there is an extended algorithm of decision tree named gradient boosted decision tree (GBDT) \cite{friedman2001greedy}\cite{chen2016xgboost}.

In this paper, we choose GBDT as our learning algorithm for three main reasons:
\begin{enumerate}
\item It does not require the input feature normalization since the decision tree is a recursive partitioning based algorithm, which reduces the overhead the feature preprocess in runtime.
\item Among 10 popular supervised learning algorithms, boosted decision tree outperforms other algorithms, including SVM and traditional decision tree on a variety of tested data sets \cite{caruana2006empirical}.
\item The prediction time complexity is acceptable, say $O(h)$, where $h$ is the depth of the trained decision tree and can be restricted to a fixed value.
\end{enumerate}
There are several algorithms of tree decision learning (e.g., ID3 \cite{quinlan1986induction}, C4.5 \cite{quinlan2014c4} and CART \cite{loh2011classification}), and CART would be more competitive in some cases compared to others \cite{anyanwu2009comparative}. So we choose CART as our model training algorithm, and we use the implementation of gradient boosting framework named XGBoost \cite{chen2016xgboost}, which is flexible, portable and highly efficient.

\textbf{Parameter Configuration.} We need to consider two main impacts when setting the parameters. On one hand, it is crucial that the depth of the decision tree should not be too deep, otherwise it will increase the overhead of the predictor in runtime. On the other hand, we need to set the proper parameters such that the prediction accuracy is high enough. In this paper, we set the maximum depth of the decision tree to be 8 and the number of estimators for boosting is also 8. We set step size shrinkage ($eta$) to be 1, and the minimum loss reduction ($gamma$) to 0, which makes the boosting algorithm more progressive.

\textbf{Training.} Instead of training model separately from different GPUs, we hope that the model is equipped with robustness to different GPU hardware, so we put all the input feature (8-dimension vector, including 5 characteristics of GPU) into one model training. We randomly split the data set into training data set (80\%) and testing data set (20\%). Note that in the 80\% training data set, there include 80\% samples from each GPU, and the remainder is used as testing data set. To validate whether the chosen model can generalize our data set, 5-fold cross-validation is presented in this work. After the evaluation of cross-validation, the whole data set is used as training data to learn the final model that can be put into real-world applications.

\textbf{Integration.} We use the learned model as our predictor of the selection system to choose the better algorithm between NT and TNN. After the model has been well trained, the final algorithm in calculating $C=A\times B^T$ is derived, and we call it MTNN. The pseudo-code of MTNN is shown in Algorithm \ref{algo:mtnn}.

\begin{algorithm}
\caption{MTNN} \label{algo:mtnn}
\begin{algorithmic}[1]
\Procedure{MTNN}{A, B, C, m, n, k}
\footnotesize{
\State // Get GPU property, O(1) complexity
\State cudaDeviceProp gpuProp;
\State cudaGetDeviceProperties(\&gpuProp, devID);
\State // Predict best algorithm
\State int label = globalLoadedPredictor(gpuProp, m, n, k);
\State // Select the algorithm: NT (1) or TNN (-1)
\If {label == 1}
  \State // Call NT of cuBLAS
\State \begin{varwidth}[t]{\linewidth}
cublasSgemm(\par
\hskip\algorithmicindent cublasHandler, \par
\hskip\algorithmicindent CUBLAS\_OP\_N, \par
\hskip\algorithmicindent CUBLAS\_OP\_T, \par
\hskip\algorithmicindent A, lda, \par
\hskip\algorithmicindent B, ldb, \par
\hskip\algorithmicindent C, ldc, ...);
\end{varwidth}
\Else
  \State // Call TNN procedure
  \State TNN(A, B, C, m, n, k);
\EndIf
\EndProcedure
}
\end{algorithmic}
\end{algorithm}

\section{Evaluation} \label{results}
We first demonstrate the evaluation of the accuracy of the predictor, which figures out the performance of the classifier, and then we present the overall performance improvement with the trained predictor (i.e., the performance of MTNN), which displays how well the selection system is.

\subsection{Performance of Classification}
To evaluate the performance of the classification algorithm, we use the metric of classifying accuracy to measure the classifiers. The average accuracy of our pre-defined 5-fold cross-validation is 90.51\%, which means that the predictor makes the calculation of $C=A\times B^T$ fast enough in 90.51\% cases. Since the testing data set is an imbalanced set with a larger number of negative samples than positive samples, both accuracies of the negative and the positive classes are recorded. Table \ref{table:5foldacc} shows the details of the accuracy of the 5-fold cross-validation.

\begin{table}[!ht]
\centering
\caption{Accuracies of the 5-fold cross-validation}
\label{table:5foldacc}
\begin{tabular}{|l|c|c|c|c|}
\hline
Class &  Minimum & Maximum & Average \\\cline{1-4}
\hline
\hline
Negative & 91.36\% & 93.30\% & 92.05\% \\\cline{1-4}
Positive & 86.49\% & 92.31\% & 88.39\%\\\cline{1-4}
Total & 89.40\% & 91.94\% & 90.51\%\\\cline{1-4}
\end{tabular}
\end{table}

We also make a comparison with SVM algorithms, including axial basis function kernel (SVM-RBF) and polynomial kernel (SVM-Poly), both of which are commonly used in supervised machine learning algorithms. We use libSVM \cite{cc01a} as SVM implementation, which is a widely used tool. The parameters for SVM are: $C=1000.0$ and $gamma=0.01$, and the input feature is normalized to the range of (0, 1). The learning algorithm of traditional decision tree (DT) is also included into the comparison to show GBDT has a better performance in terms of accuracy and running efficiency. In the tested experimental environment (Table \ref{table:hardwareclf}) for learning algorithms, the performances of classifiers are shown in Table \ref{table:compwithother}.
\begin{table}[!ht]
\centering
\caption{The experimental environment for classifiers}
\label{table:hardwareclf}
\begin{tabular}{|c|c|c|c|}
\hline
CPU  &  Memory  & OS & Frequency \\\cline{1-4}
\hline
\hline
Intel CPU i7-3820 & 64 GB & Ubuntu 14.04 & 3.6 GHz\\\cline{1-4}
\end{tabular}
\end{table}

\begin{table}[!ht]
\centering
\caption{Comparison with SVM and DT}
\label{table:compwithother}
\begin{tabular}{|l|c|c|c|c|}
\hline
Classifier &  Accuracy (\%) & Train Time (ms) & Predict Time (ms) \\\cline{1-4}
\hline
\hline
GBDT & \textbf{90.51} & 7 & 0.005 \\\cline{1-4}
SVM-RBF & 81.66 & 47 & 1.2 \\\cline{1-4}
SVM-Poly & 77.68 & 30 & 1.07 \\\cline{1-4}
DT & 87.84 & \textbf{1} & \textbf{0.004} \\\cline{1-4}
\end{tabular}
\end{table}
From Table \ref{table:compwithother}, in terms of the prediction accuracy, GBDT is much better than both SVMs and DT. Regarding the training and prediction efficiency, GBDT outperforms both two types of SVMs. Even though the prediction time of GBDT is slightly longer than that of DT, it could be neglectable (only 0.005 ms) compared with the computation time of matrix-matrix multiplication.

\begin{figure}[!ht]
  \centering
  \includegraphics[width=0.5\linewidth]{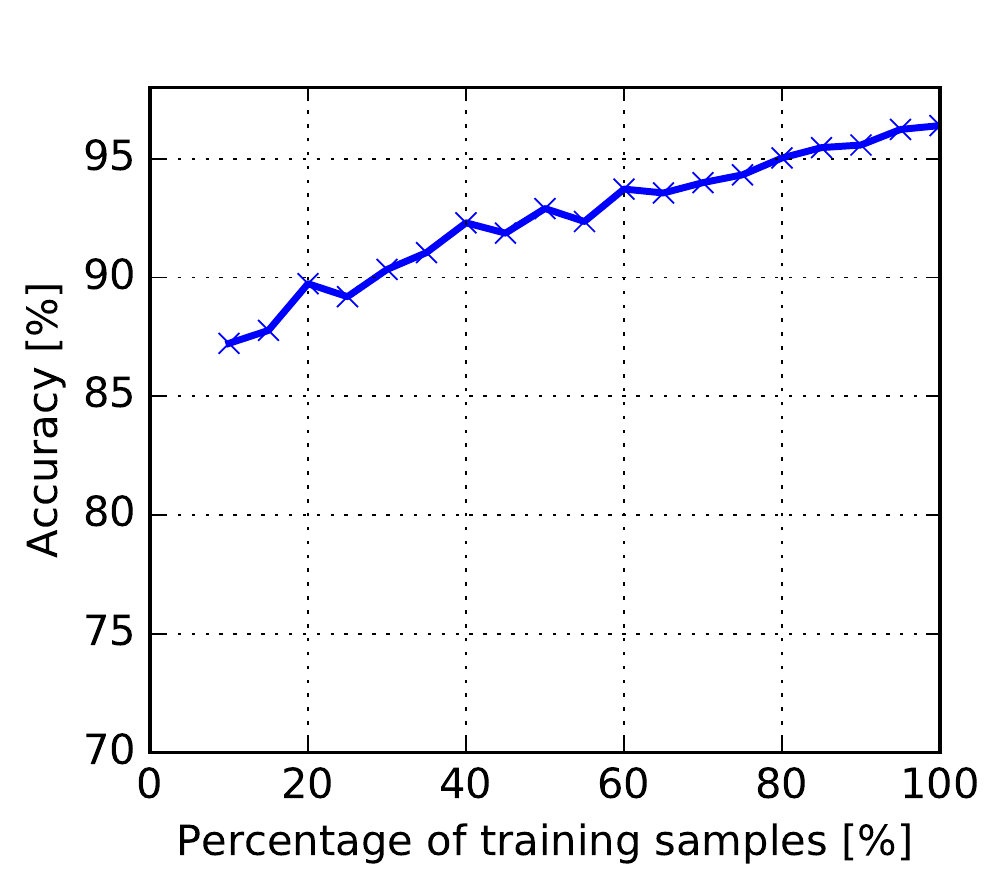}\\
\caption{Training accuracy with different size of training data sets}
\label{fig:xgconvergence}
\end{figure}

Before putting the model into the MTNN algorithm, high-accuracy model should be trained with specific parameters and training samples. The 5-fold cross validation has verified our model is admissible with 80\% training samples, but there exists a question that how many training samples should be chosen for the better convergence of the model so that MTNN has a higher prediction accuracy. We use different sizes of the training data set to figure out how many samples are proper to train a high-accuracy predictor. From all the 1832 samples, $x$ percent are selected as the training data set, and the whole samples are used as the testing data set, where $x$ is selected from 10 to 100 with a step size of 5. The training accuracy with different size of the training data set is shown in Fig. \ref{fig:xgconvergence}. It displays a tend of higher accuracy with larger size of training data set.
\subsection{Performance of Selection}
\begin{figure*}[!ht]
  \centering
  \includegraphics[width=\linewidth]{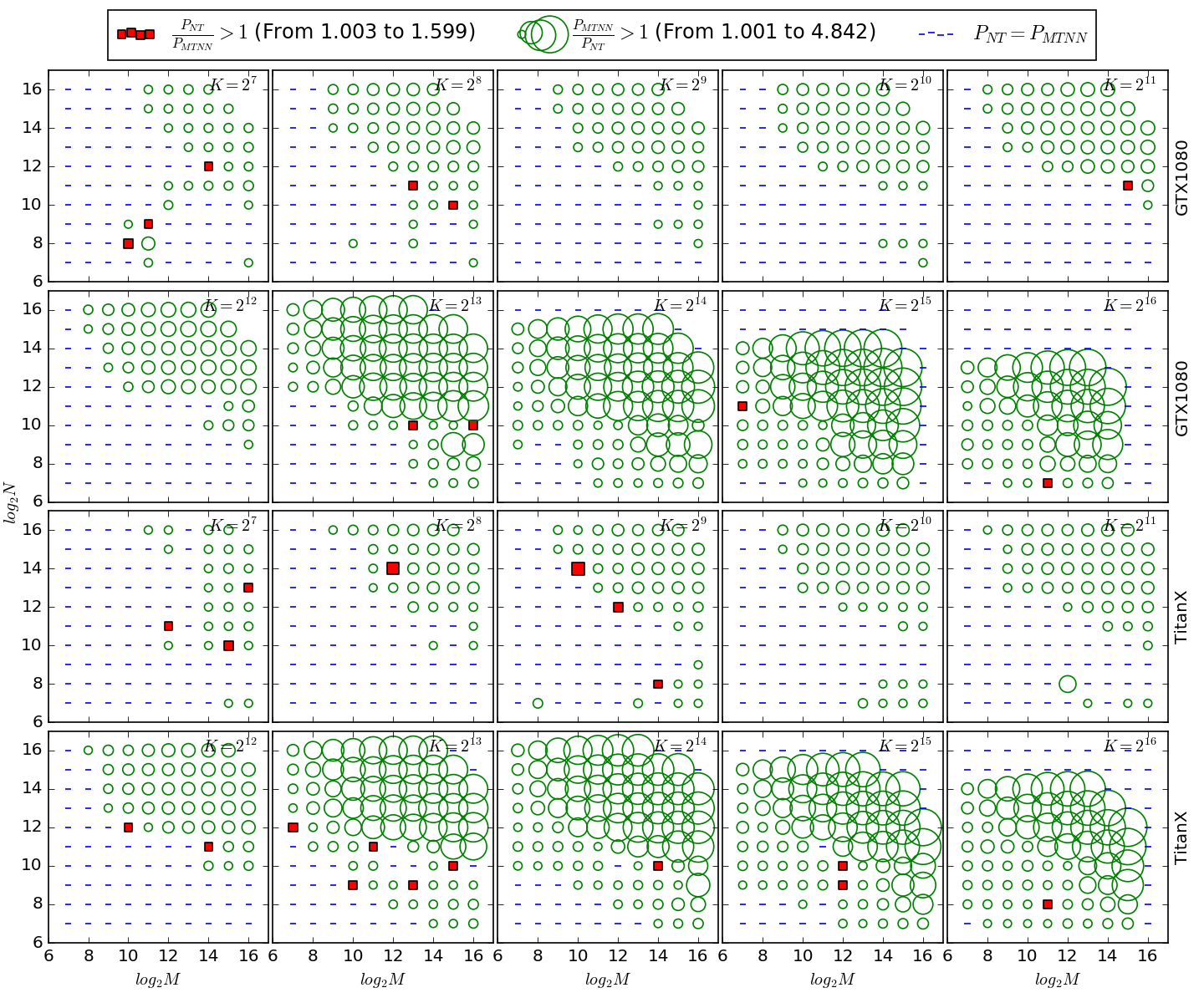}\\
\caption{The performance comparison between NT and MTNN method in calculating $C=A\times B^T$. The rectangle symbol in the legend indicates that the performance of NT is better than MTNN, and the circle symbol green color indicates that the performance of NT is worse than MTNN, and the dash symbol with blue color indicates that the performances of NT and MTNN are equal.}
\label{fig:ntvsmtnn}
\end{figure*}

In this section, we want to show that how much performance improved by using the MTNN algorithm, which is integrated with the trained predictor. In the algorithm of MTNN, the integrated predictor is trained with all the data set to achieve higher performance instead of just using 80\% data for training because the more data the higher accuracy in general. As we can see from Fig. \ref{fig:xgconvergence}, with 100\% data as training set, the trained predictor with GBDT achieves 96.39\% accuracy in classification, which means the selection system makes the correct decision to choose the better algorithm between NT and TNN in 96.39\% cases.

\begin{figure}[!htbp]
  \centering
  \subfigure[GTX1080]
  {
    \includegraphics[width=0.45\linewidth]{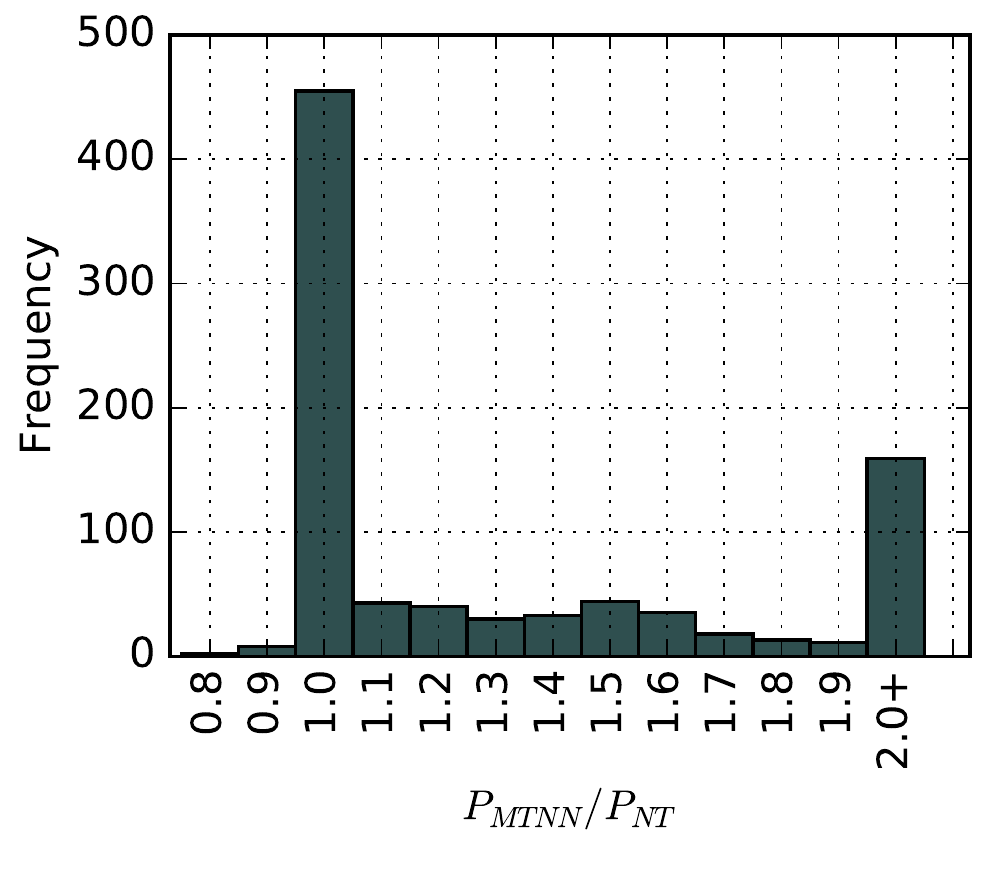}
  }
  \subfigure[TitanX]
  {
    \includegraphics[width=0.45\linewidth]{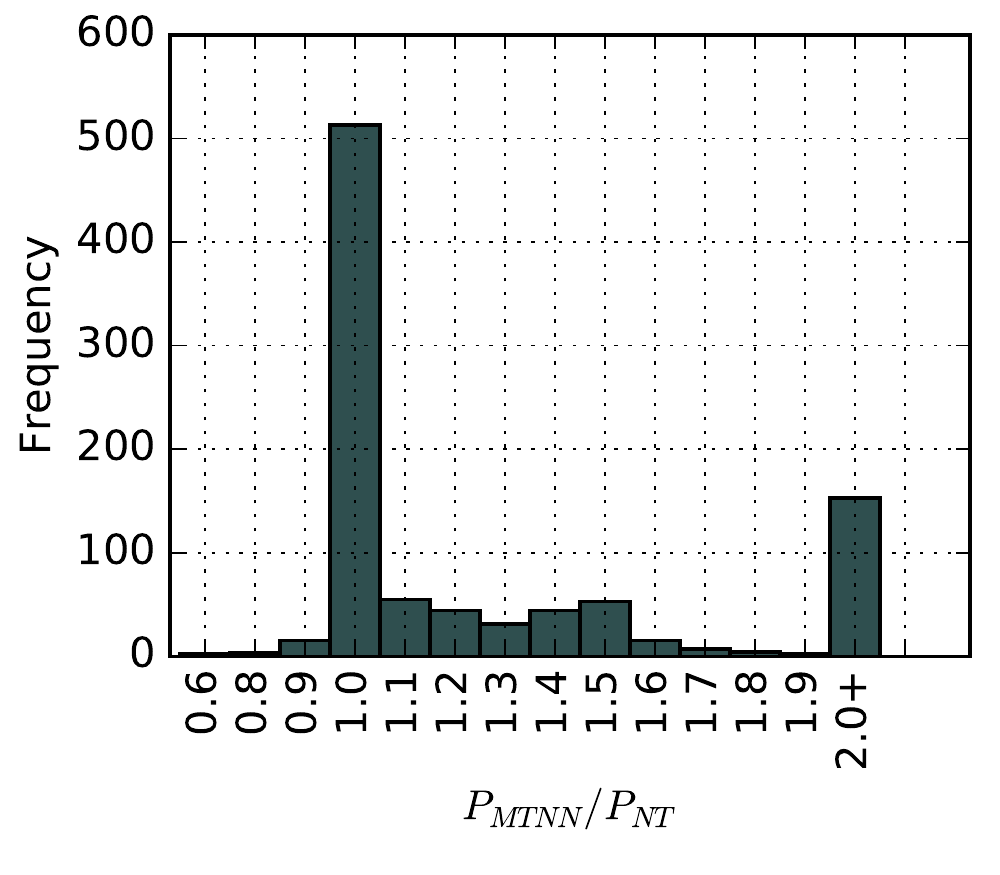}
  }
\caption{The frequency of performance ratio of $P_{MTNN}$ over $P_{NT}$ among tested cases on each GPU. The last value (e.g, 2.0+) of x-axis means: $P_{MTNN}/P_{NT}$ is greater than or equal to that value (e.g., $P_{MTNN}/P_{NT} \geq 2.0$).}
\label{fig:ntvsmtnnfreq}
\end{figure}

Before presenting the statistic results of MTNN compared to NT and TNN, a visualized comparison between MTNN and NT on our tested GPUs is shown in Fig. \ref{fig:ntvsmtnn}. Compared to Fig. \ref{fig:ntvstnn}, the red rectangles, which indicate that the performance of TNN is worse than NT, are reduced to a very small portion by the MTNN method. In other words, in most cases, the performance of MTNN is better than or equal to NT; and only in a minority of cases, the performance of MTNN is worse than NT. The statistic frequency on the performance of MTNN over NT is shown in Fig. \ref{fig:ntvsmtnnfreq}. The portion of the cases that MTNN outperforms NT is 47.81\% on GTX1080, and 43.35\% on TitanX. It shows that there is futher optimization space for the matrix-matrix-transpose multiplication algorithm on Pascal GPUs. In Fig. \ref{fig:ntvstnn}, the maximum value of $P_{NT}/P_{TNN}$ is 15.394, while Fig. \ref{fig:ntvsmtnn} displays the maximum of $P_{NT}/P_{MTNN}$ is only about 1.6.

Similar to the work in \cite{spillinger2015matrix} and to make further comparisons in a statistic way, we use $GOW$ (Gain over Worst) to denote \textbf{G}ain in performance of MTNN \textbf{O}ver the \textbf{W}orst algorithm at each sample. $GOW$ is calculated by:
\begin{equation}
GOW = \frac{P_{MTNN} - min(P_{NT}, P_{TNN})}{min(P_{NT}, P_{TNN})}
\end{equation}
Let $LUB$ (Loss under Best) denote the percent \textbf{L}oss of MTNN \textbf{U}nder the \textbf{B}est algorithm for each sample, which is calculated by:
\begin{equation}
LUB = \frac{P_{MTNN} - max(P_{NT}, P_{TNN})}{max(P_{NT}, P_{TNN})}
\end{equation}
We can define some metrics to measure the performance of MTNN compared to NT and TNN. The description of metrics is displayed in Table \ref{table:metricinfo}. And the corresponding evaluated values are shown in Table \ref{table:metriceval}.

\begin{table}[!ht]
\centering
\caption{Metrics description}
\label{table:metricinfo}
\begin{tabular}{|l|l|l|l|l|}
\hline
Metric &  Description \\\cline{1-2}
\hline
\hline
\textit{MTNN vs NT} & Average percent improvement of using MTNN versus \\
    & versus always choosing TN\\\cline{1-2}
\textit{MTNN vs TNN} & Average percent improvement of using MTNN versus \\
    & versus always choosing TNN\\\cline{1-2}
$GOW_{avg}$ & Average $GOW$ in all samples \\\cline{1-2}
$GOW_{max}$ & Maximum $GOW$ in all samples \\\cline{1-2}
$LUB_{avg}$ & Average $LUB$ in all samples \\\cline{1-2}
$LUB_{min}$ & Maximum $LUB$ in all samples \\\cline{1-2}
\end{tabular}
\end{table}

\begin{table}[!ht]
\centering
\caption{Values of performance metrics of MTNN in \%}
\label{table:metriceval}
\begin{tabular}{|l|c|c|c|}
\hline
Metric & GTX1080 & TitanX & Total   \\\cline{1-4}
\hline
\hline
\textit{MTNN vs NT}   &57.78      &50.48        & 54.03 \\\cline{1-4}
\textit{MTNN vs TNN}  &21.51      &22.31        & 21.92 \\\cline{1-4}
$GOW_{avg}$           &79.44      &73.20        & 76.23  \\\cline{1-4}
$GOW_{max}$           &1439.39    &957.44       & 1439.39  \\\cline{1-4}
$LUB_{avg}$           &-0.15      &-0.40        & -0.28  \\\cline{1-4}
$LUB_{min}$           &-25.07     &-71.62       & -71.62  \\\cline{1-4}
\end{tabular}
\end{table}

From Table \ref{table:metriceval}, we can see MTNN achieves 54.03\% performance improvement compared to use the NT algorithm only, and 21.92\% compared to TNN on average. Compared to the worst cases of NT and TNN, MTNN achieves up to 76.23\% performance improvement on average and up to 1439.39\% in some particalar cases. There are some cases that the predictor makes the wrong decision, but the slowdown performance is only about 0.28\%. In other words, compared to the best cases of NT and TNN, the performance of MTNN is only 0.28\% worse when the predictor chooses the lower performance algorithm of NT and TNN. Between these two GPUs, the speedup of time efficiency on the GTX1080 card is slightly higher than that on the TitanX card.

\subsection{Evaluation with Caffe}
\begin{figure}[!ht]
  \centering     
  \subfigure[2 hidden layers on GTX1080]
  {
    \includegraphics[width=0.45\linewidth]{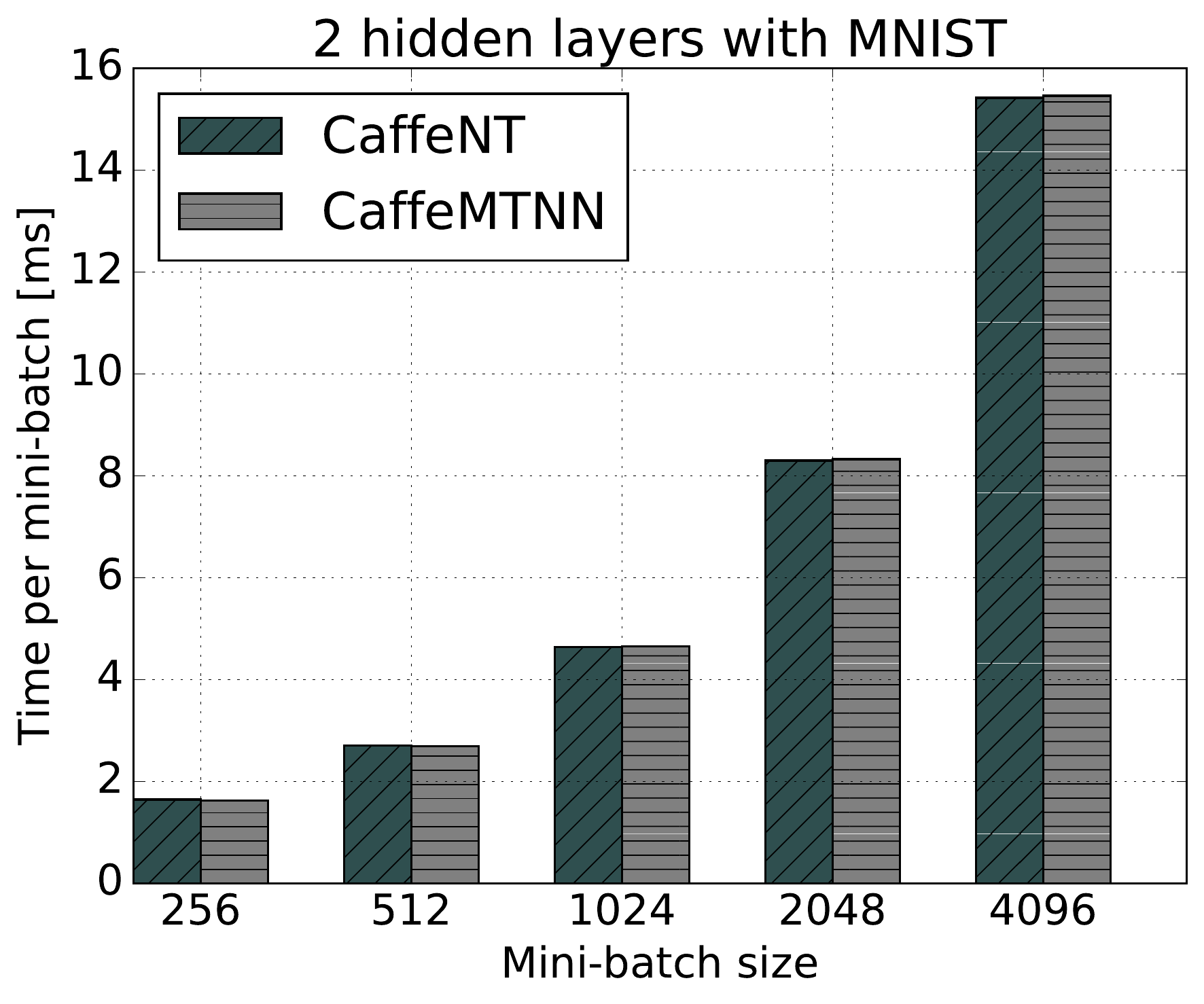}
  }
  \subfigure[2 hidden layers on TitanX]
  {
    \includegraphics[width=0.45\linewidth]{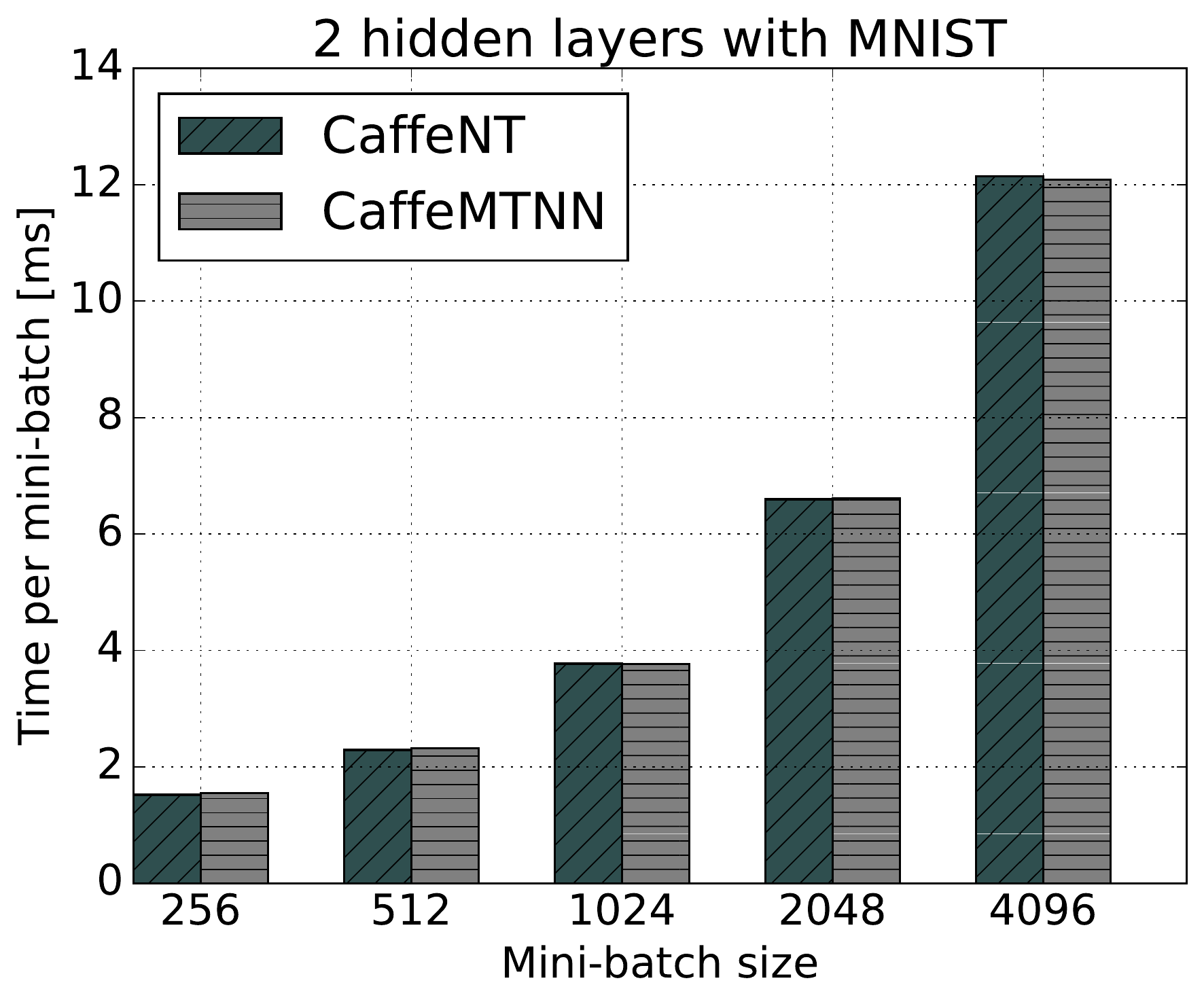}
  }
  \subfigure[3 hidden layers on GTX1080]
  {
    \includegraphics[width=0.45\linewidth]{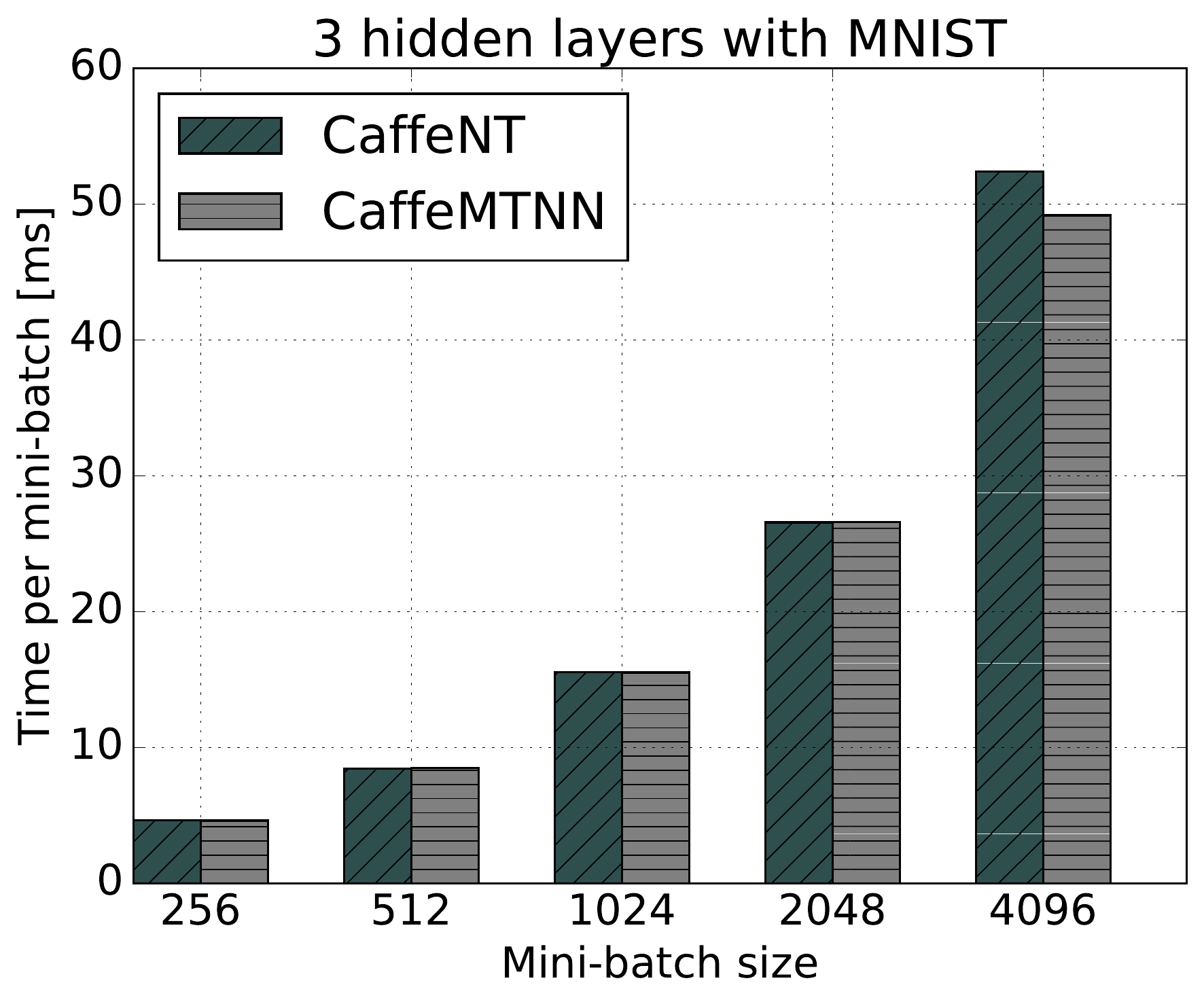}
  }
  \subfigure[3 hidden layers on TitanX]
  {
    \includegraphics[width=0.45\linewidth]{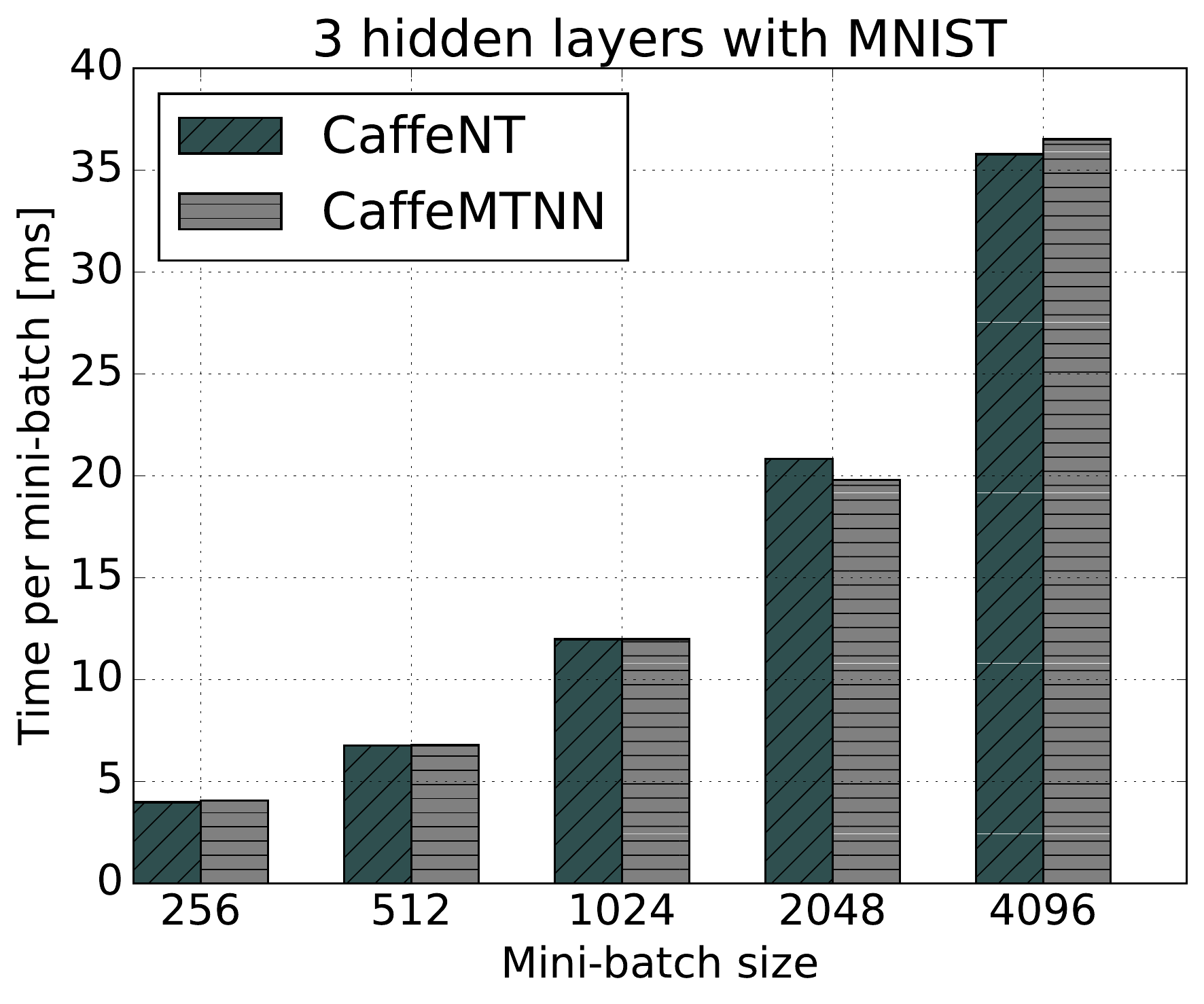}
  }
  \subfigure[4 hidden layers on GTX1080]
  {
    \includegraphics[width=0.45\linewidth]{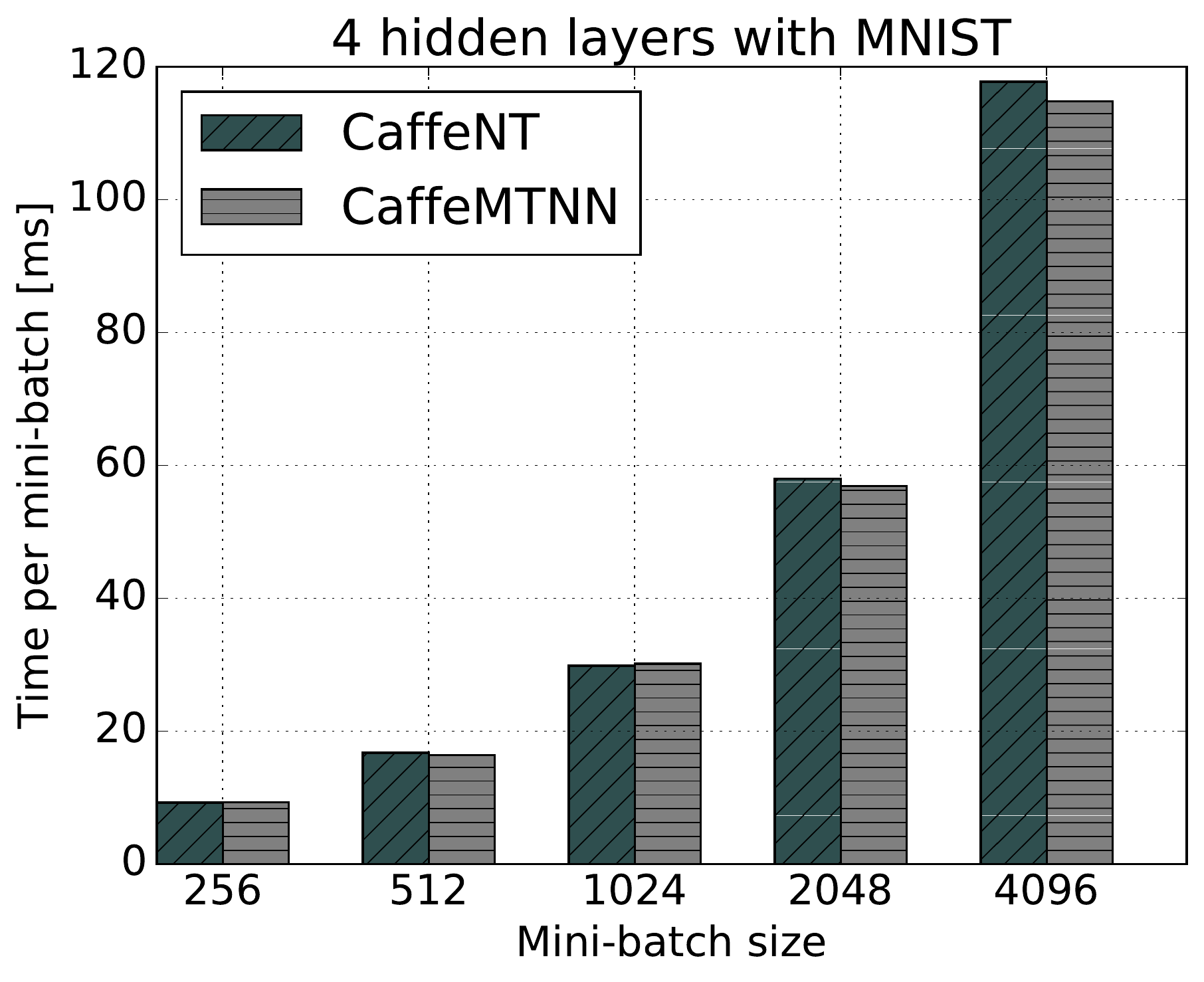}
  }
  \subfigure[4 hidden layers on TitanX]
  {
    \includegraphics[width=0.45\linewidth]{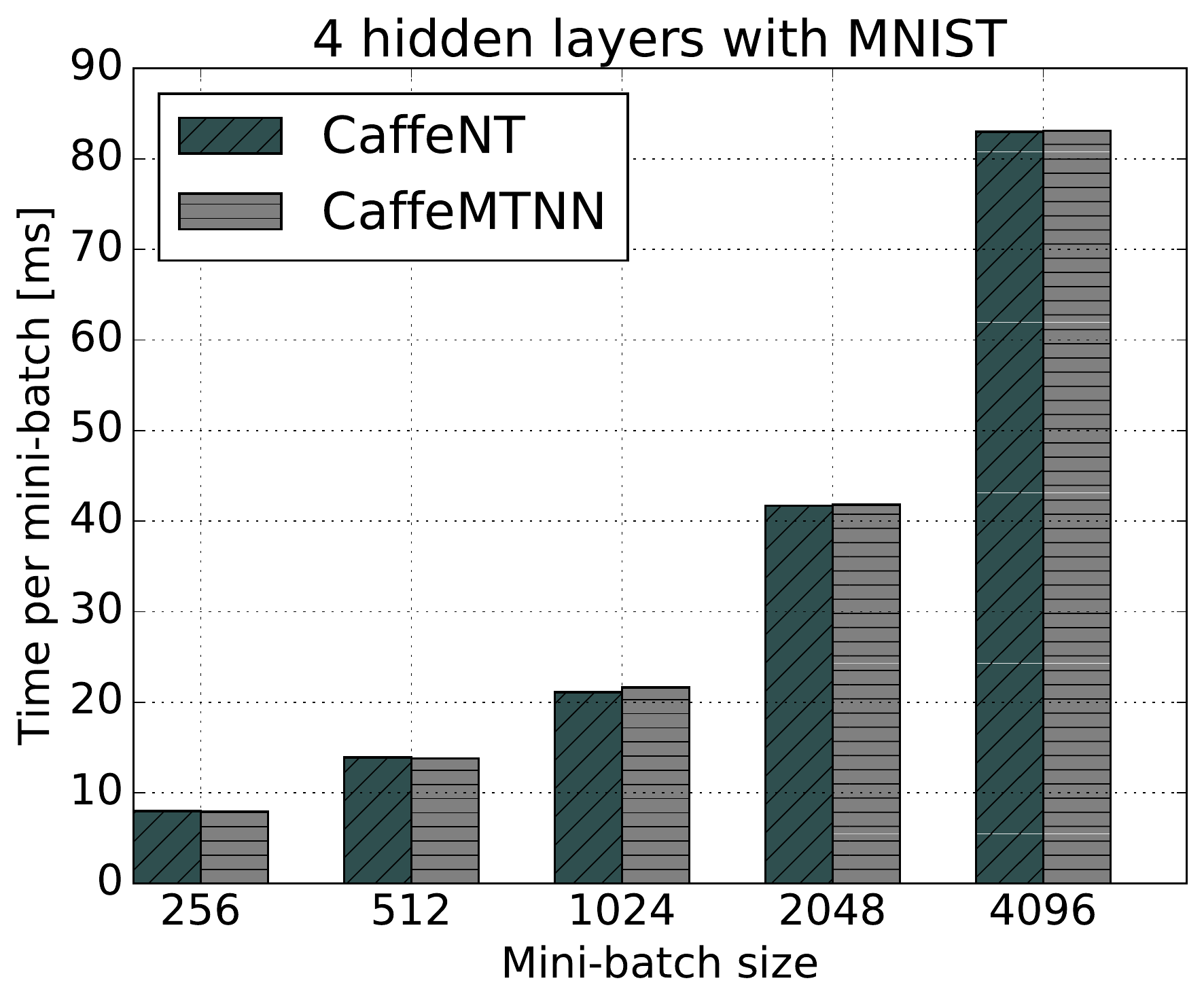}
  }
  \caption{The performance comparison with MNIST between the original Caffe and our optimzed Caffe (The lower the better).}
  \label{fig:caffevsours_gpu_mn}
\end{figure}

\begin{figure}[!ht]
  \centering     
  \subfigure[2 hidden layers on GTX1080]
  {
    \includegraphics[width=0.45\linewidth]{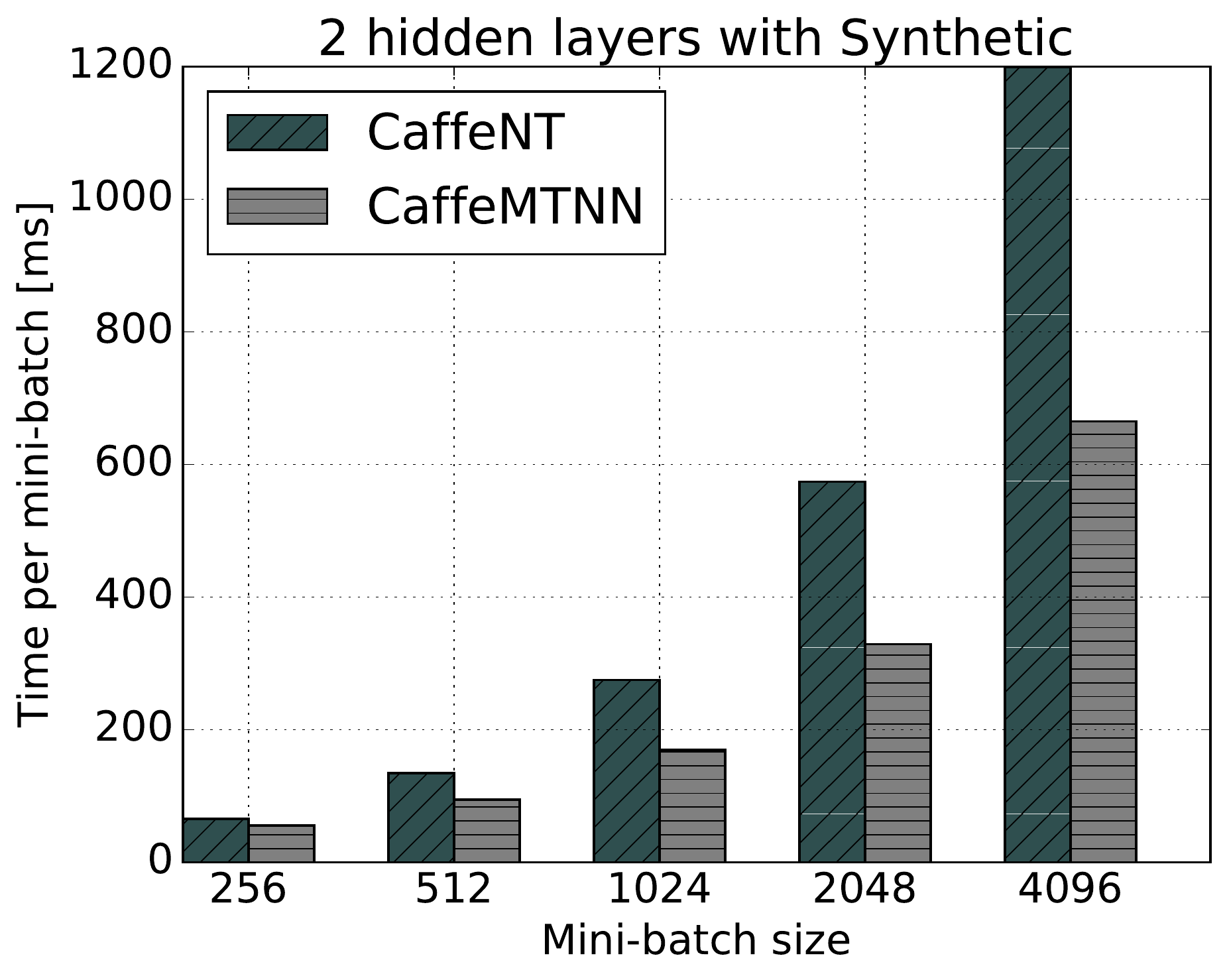}
  }
  \subfigure[2 hidden layers on TitanX]
  {
    \includegraphics[width=0.45\linewidth]{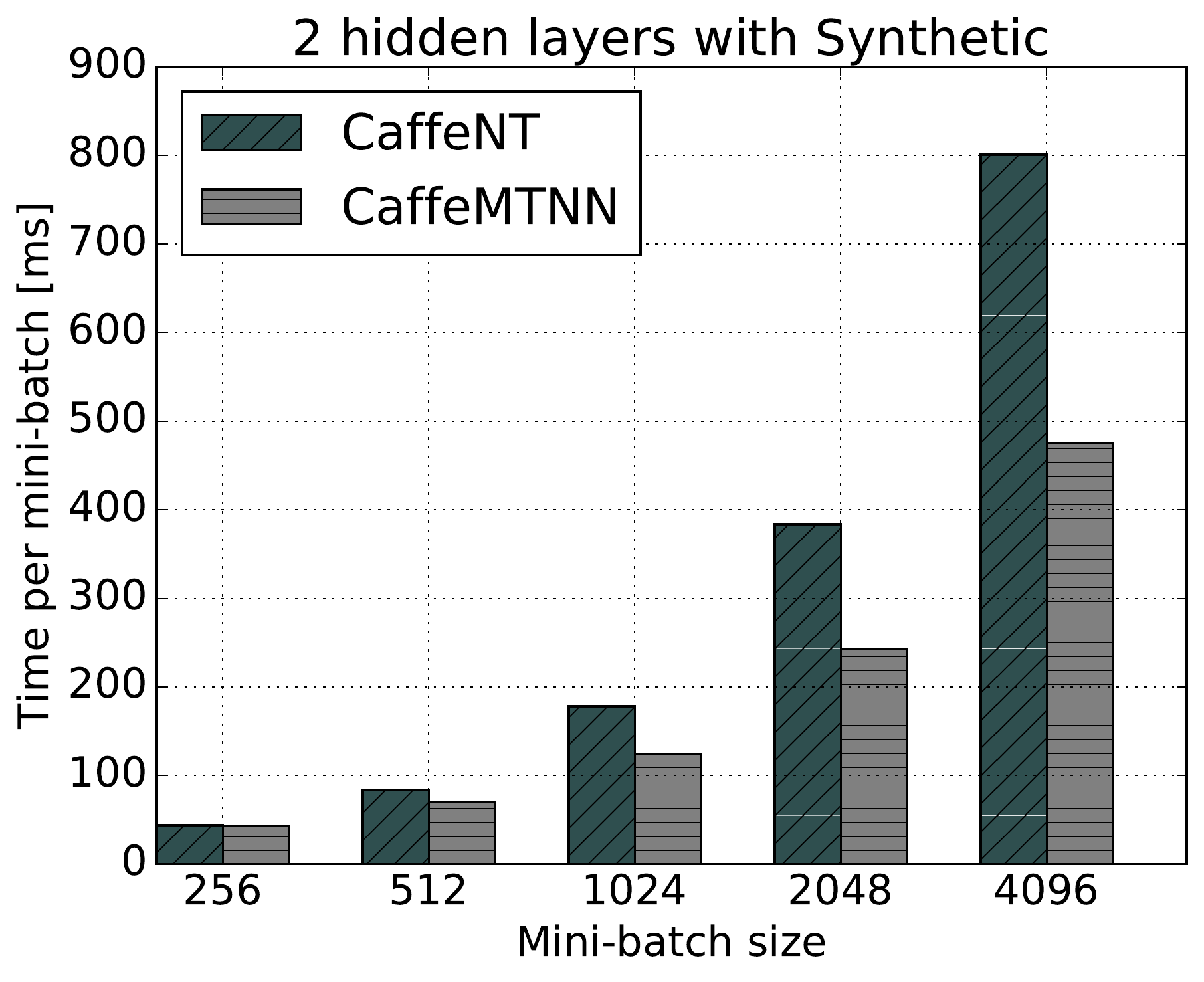}
  }
  \subfigure[3 hidden layers on GTX1080]
  {
    \includegraphics[width=0.45\linewidth]{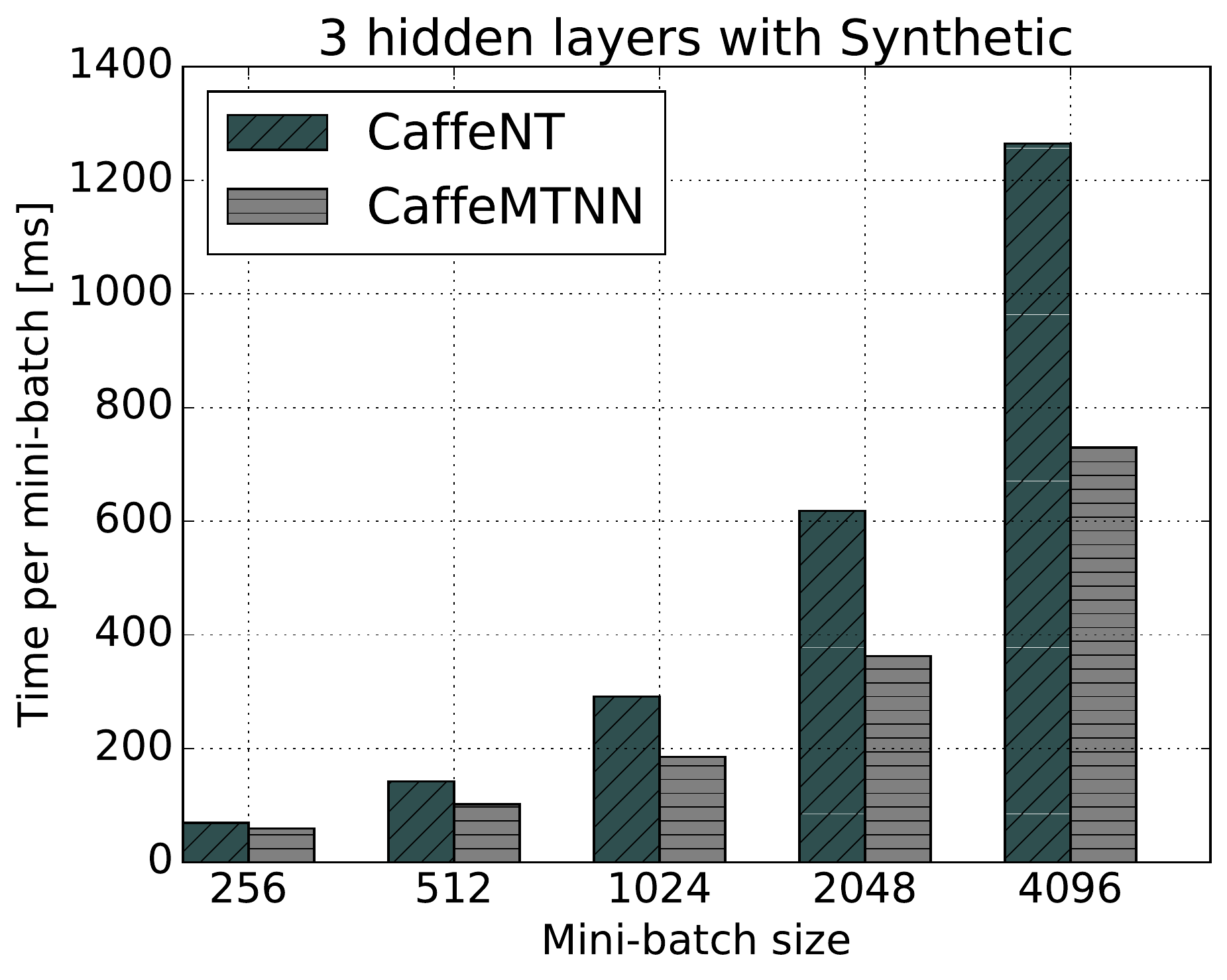}
  }
  \subfigure[3 hidden layers on TitanX]
  {
    \includegraphics[width=0.45\linewidth]{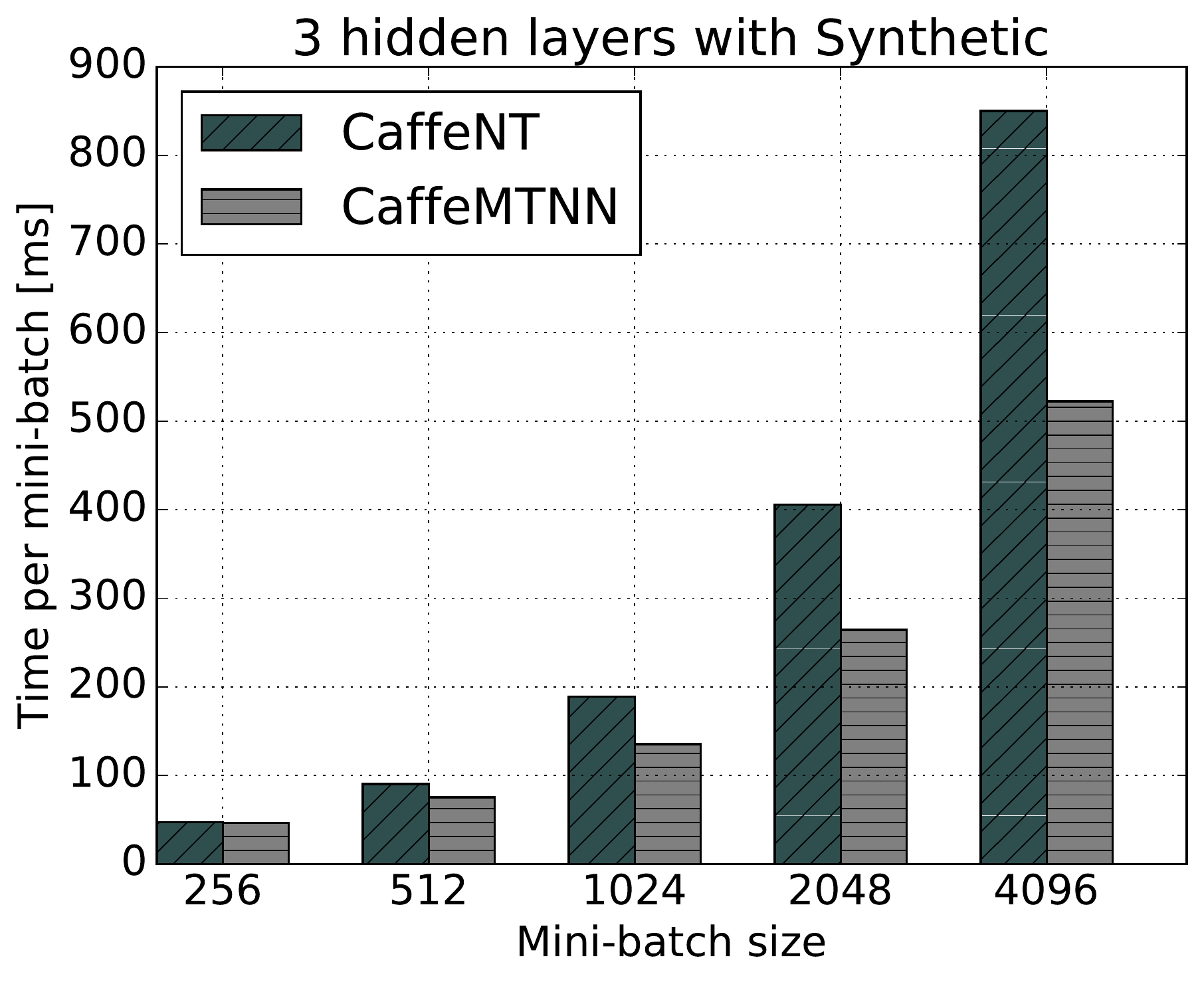}
  }
  \subfigure[4 hidden layers on GTX1080]
  {
    \includegraphics[width=0.45\linewidth]{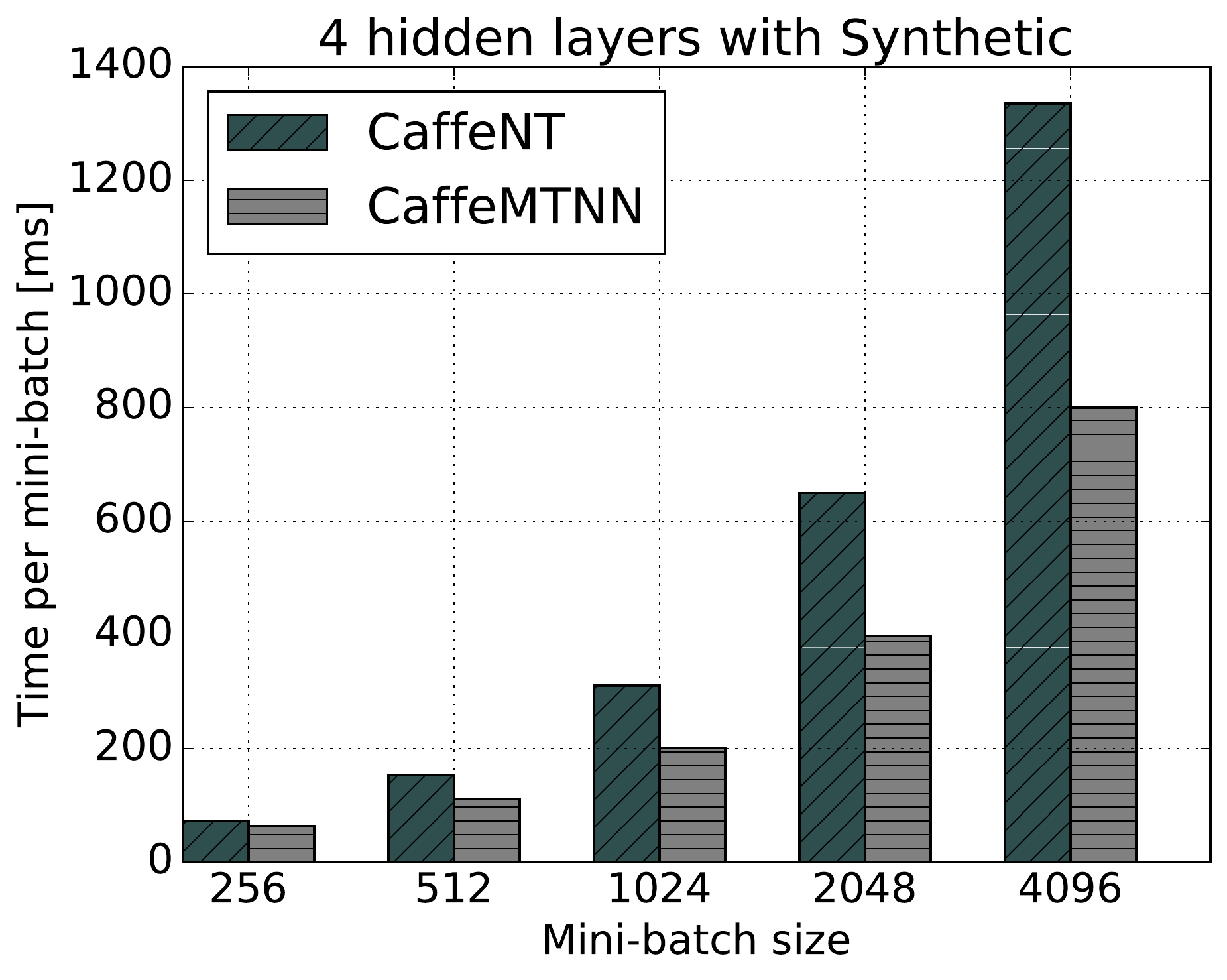}
  }
  \subfigure[4 hidden layers on TitanX]
  {
    \includegraphics[width=0.45\linewidth]{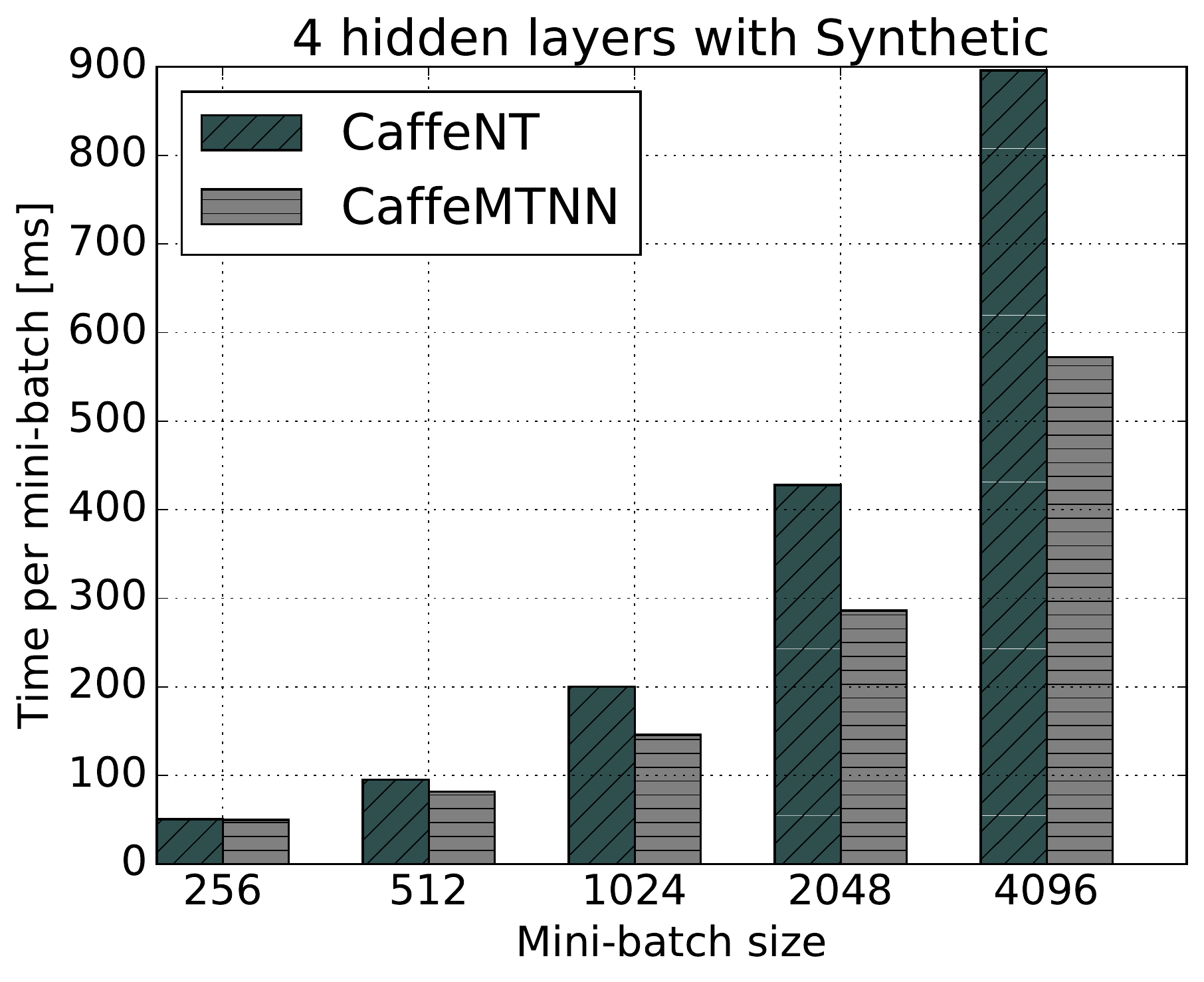}
  }
  \caption{The performance comparison on larger FCN between the original Caffe and our optimzed Caffe (The lower the better).}
  \label{fig:caffevsours_gpu_sy}
\end{figure}

To test the performance of MTNN in the real-world application, we integrate the MTNN algorithm into Caffe \cite{jia2014caffe} which is one of the most popular deep learning frameworks. We choose two types of fully connected networks: one is with the MNIST data set whose input and output dimensions are small, and the other one is with a synthetic data whose input and output dimensions are large. For each type of fully connected network, a variety of hidden layers are configured, namely 2, 3 and 4 layers. The configuration details of neural networks are shown in Table \ref{table:fcnconf}. The performance comparison of these two types of networks running on the original version of Caffe (CaffeNT) and Caffe with MTNN (CaffeMTNN), are displayed in Fig. \ref{fig:caffevsours_gpu_mn} and Fig. \ref{fig:caffevsours_gpu_sy}, respectively.

\begin{table}[!ht]
\centering
\caption{Fully connected networks configuration for evaluation}
\label{table:fcnconf}
\begin{tabular}{|l|l|l|l|}
\hline
Data set & MNIST & Synthetic   \\\cline{1-3}
\hline
\hline
Input & 784 & 26752 \\\cline{1-3}
Output & 10 & 26752 \\\cline{1-3}
2 hidden layers & 2048-1024 & 4096-4096 \\\cline{1-3}
3 hidden layers & 2048-2048-1024 & 4096-4096-4096 \\\cline{1-3}
4 hidden layers & 2048-2048-2048-1024 & 4096-4096-4096-4096 \\\cline{1-3}
\end{tabular}
\end{table}

By integrating our method to Caffe, the performance of the optimized Caffe accomplishes a slightly improvement of 1.74\% with the MNIST data set, while the performance improvement is as much as 28.2\% with the synthetic data set.

On one hand, from Fig. \ref{fig:caffevsours_gpu_mn}, it is noted that the training time speed of CaffeNT and CaffeMTNN is very close with all the mini-batch sizes. The main reason is that with specific number of neurons in two adjacent layers (e.g., $l1$ and $l2$) and the mini-batch size ($mb$), the size of matrix-matrix-transpose multiplication is decided by $l1$, $l2$ and $mb$. If the values of $l1$, $l2$ and $mb$ are too small, the performance of TNN has no advantages compared to the original NT of cuBLAS, which can be explained with the performance comparison in Fig. \ref{fig:ntvsmtnn} (there are many dash symbols on the left-bottom side of the figure, so MTNN can only be on the par with NT of cuBLAS). There exists a particular case that MTNN is slightly worse than NT of cuBLAS with mini-batch of 4096 in the network of 3 hidden layers on TitanX. The reason of this minor slowdown is that the predictor makes the error prediction, but it may occur only in a very small probability since the accuracy of the predictor is up to 96\%.

On the other hand, from Fig. \ref{fig:caffevsours_gpu_sy}, with the larger neural network (the input size and the output size are both 27652 in our tested case) and the larger mini-batch size (larger than 512), the speedup of CaffeMTNN is significant. And the matrix-matrix-transpose multiplication can be mapped to the cases in the right-top side of Fig. \ref{fig:ntvsmtnn}, where it has numerous green circles, which means the deep neural networks can benefit from the higher performance algorithm of MTNN. 
\begin{table}[!ht]
\centering
\caption{Breakdown of the average running time in millisecond and speedups}
\label{table:breakdownresults}
\begin{tabular}{|l|l|l|c|c|c|}
\hline
Data set& GPU & Phase & CaffeNT & CaffeMTNN & Speedup \\\cline{1-6}
\hline
\hline
	  &		&Forward  	&11.15 & 10.39 & \textbf{1.07}		\\\cline{3-6}
	 &G.1080&Backward 	&58.81 & 59.79 & 0.98		\\\cline{3-6}
MNIST &		&Total 		&24.79 & 24.31 & 1.02		\\\cline{2-6}\cline{2-6}

	 &		&Forward 	&7.36 & 7.38 & 1.00			\\\cline{3-6}
	 &TitanX&Backward 	&47.69 & 47.39 & 1.01		\\\cline{3-6}
	 &		&Total 		&18.22 & 18.25 & 1.00		\\\cline{1-6}\cline{1-6}

	  &		&Forward 	&320.83 & 131.62 & \textbf{2.44}		\\\cline{3-6}
	 &G.1080&Backward 	&1029.77 & 1033.04 & 1.00	\\\cline{3-6}
Synth- &		&Total 	&477.05 & 288.24 & 1.66 	\\\cline{2-6}\cline{2-6}

etic &		&Forward 	&200.54 & 93.12 & \textbf{2.15}	\\\cline{3-6}
	 &TitanX&Backward 	&761.08 & 763.59 & 1.00	\\\cline{3-6}
	 &		&Total 		&316.13 & 208.99 & 1.51	\\\cline{1-6}
\end{tabular}
\end{table}
The matrix-matrix-tranpose multiplication only impacts either the forward propagation or the backward propagation during the training of deep neural networks. To demonstrate which phase benifits from the MTNN method, we break down the running time in one mini-batch to the forward phase and the backward phase in the experimental results. Instead of showing all the tested cases seperately, we show the statistic results with different data sets on different GPUs by averaging all the mini-batch sizes and layers. The results are shown in Table \ref{table:breakdownresults}. It is noted that the running time of the backward propagation is almost the same in all the cases. The main speedup of the training process is contributed to the forward phase. With the MNIST data set whose network size is small, CaffeMTNN is on the par with CaffeNT. With the sythetic data set whose network size is large, the speedup of the forward propagation of CaffeMTNN is significant, and it obtains as much as 2.44x and 2.15x speedups compared to CaffeNT on GTX1080 and TitanX, respectively.

\section{Conclusion and Future Work} \label{conclusion}
In this paper, we first figure out the low performance of cuBLAS in calculating the matrix-matrix-transpose multiplication compared to the matrix-matrix multiplication by benchmarking a variety of cases. To accelerate the calculation of matrix-matrix-transpose multiplication, we propose a simple solution (named TNN), which carrys out the efficient out-of-place tranpose algorithm first and then make use of the high performance matrix-matrix multiplication algorithm. TNN achieves some performance improvement, but it still may fall into even worse efficiency. In order to obtain the best average performance, we design a supervised learning based algorithm (named MTNN), which can make an intelligent choose of proper algorithm in calculating matrix-matrix-transpose multiplication. Using the boost gradient decision tree learning algorithm, MTNN can carry out the matrix-matrix-transpose multiplication with faster routine in an accuracy of 96\%. We evaluate the performance of our algorithm on two modern GPUs (GTX1080 and Titan X Pascal). The experimental results show that the MTNN method achieves 54.03\% performance improvement compared to cuBLAS. To verify the effectiveness of MTNN in the real-world application, we integrate MTNN into a popular deep learning framework: Caffe, and the optimized Caffe obtains an average of 28\% improvement on fully connected networks.

The transpose algorithm used in this paper is an out-of-place method, which requires extra memory to store the transpose of matrix. The selection system could not be used if the GPU card has no enough memory. Therefore, we plan to exploit in-place matrix transpose algorithms by finding a good trade-off between the memory overhead and throughput.

\bibliographystyle{IEEEtran}
\Urlmuskip=0mu plus 1mu
\bibliography{mmv5-r1.bbl}

\end{document}